\documentclass[a4paper,12pt]{article}

\usepackage{amsmath}
\usepackage{enumerate}
\usepackage{natbib}
\usepackage{url}

\usepackage{amsfonts}
\usepackage{bm}
\usepackage{adjustbox}
\usepackage{appendix}

\usepackage{color}
\usepackage{colortbl}
\usepackage{graphics}
\usepackage{wrapfig}
\usepackage{caption}
\usepackage{float}

\usepackage{afterpage}
\usepackage{threeparttable}

\RequirePackage[left=2.2cm,
right=2.2cm,
top=2.5cm,
bottom=2.5cm,headheight=12pt]{geometry}
\allowdisplaybreaks
\makeatother

\graphicspath{{Figures/}}

\usepackage{physics} % jump-diff model \diff; \abs

\newcommand{\blue}[1]{{\textcolor{darkblue}{#1}}}

\usepackage{float}

\usepackage{csquotes}

\usepackage{afterpage}
\usepackage{lscape}

\floatstyle{plaintop}
\restylefloat{table}

\usepackage[table]{xcolor}

\newcommand{\grb}[1]{\cellcolor[gray]{0.9}{#1}}
\newcommand{\grbb}[1]{\cellcolor[gray]{0.7}{#1}}

\usepackage{tikz}
\usetikzlibrary{matrix,shapes,arrows,fit}

\tikzset{
	every picture/.style={remember picture,baseline},
	every node/.style={anchor=base,align=center,outer sep=1.5pt},
	every path/.style={thick},
}

\usepackage{pifont}% http://ctan.org/pkg/pifont

\usepackage{amsthm}

\usepackage{amsmath}               
{
	\theoremstyle{plain}
	
}

{
	\theoremstyle{plain}
	
}

{
	\theoremstyle{plain}
	
}

{
	\theoremstyle{plain}
	
}

\usepackage{subfig}

\definecolor{darkblue}{rgb}{0.0, 0.0, 0.55}
\definecolor{cadmiumgreen}{rgb}{0.0, 0.42, 0.24}
\usepackage[colorlinks,citecolor=black,linkcolor=black,urlcolor=black]{hyperref}  % darkblue
\usepackage{units}

\usepackage{mathtools}

\DeclarePairedDelimiter\floor{\lfloor}{\rfloor}

\usepackage{pdflscape}

\usepackage{amssymb}

\newcommand{\comment}{0} % 0: no comments / 1: comments 

\usepackage{lipsum,atbegshi,etoolbox}% http://ctan.org/pkg/{lipsum,atbegshi,etoolbox}
\makeatletter
\newcommand{\discardpages}[1]{% \discardpages{<csv list>}
	\xdef\discard@pages{#1}% Store pages to discard
	\AtBeginShipout{% At shipout, decide whether to discard page/not
		\renewcommand*{\do}[1]{% How to handle each page entry in csv list
			\ifnum\value{page}=##1\relax%
			\AtBeginShipoutDiscard% Discard page/not
			\gdef\do####1{}% Do nothing further
			\fi%
		}%
		\expandafter\docsvlist\expandafter{\discard@pages}% Process list of pages to discard
	}%
}
\newif\ifkeeppage
\newcommand{\keeppages}[1]{% \keeppages{<csv list>}
	\xdef\keep@pages{#1}% Store pages to keep
	\AtBeginShipout{% At shipout, decide whether to discard page/not
		\keeppagefalse%
		\renewcommand*{\do}[1]{% How to handle each page entry in csv list
			\ifnum\value{page}=##1\relax%
			\keeppagetrue% Page should be kept
			\gdef\do####1{}% Do nothing further
			\fi%
		}%
		\expandafter\docsvlist\expandafter{\keep@pages}% Process list of pages to keep
		\ifkeeppage\else\AtBeginShipoutDiscard\fi% Discard page/not
	}%
}
\makeatother
\begin{document}

	\title{\bf{
			Sluggish news reactions: 
			A combinatorial approach for synchronizing stock jumps%
        \footnote{
				We have received helpful comments and suggestions from 
				Andres Algaba, Geert Dhaene, Jean-Yves Gnabo, 
                Roxana Halbleib, 
                Ilze Kalnina, 
                Nathan Lassance, Oliver Linton, André Lucas, Kristien Smedts, Lisa Van den Branden, Steven Vanduffel, Brecht Verbeken 
				and 
				the conference and 	seminar participants 
				at 
				KU Leuven, Vrije Universiteit Brussel, 
				Vrije Universiteit Amsterdam, 
				the 
				Computational and Financial Econometrics Conference (2021), the 
                Belgian Financial Research Forum (2023), the Quantitative and Financial Econometrics Conference (2023), and the Society of Financial Econometrics Summer School (2023). 
				Nabil Bouamara gratefully acknowledges  support from the Flemish Research Foundation  (FWO  fellowship \#11F8419N) and the Platform for Education and Talent (Gustave Bo\"el -- Sofina fellowship). 
				Sébastien Laurent has received fundings from the french government under the “France 2030” investment plan managed by the French National Research Agency (reference: ANR-17-EURE-0020 and ANR-21-CE26-0007-01) and from Excellence Initiative of Aix-Marseille University - A*MIDEX.
			}
	}}
	
	\author{
		Nabil Bouamara	\\
        Department of Finance, Université catholique de Louvain
        \and
        Kris Boudt\\
        Department of Economics, Ghent University  \\
        Solvay Business School, Vrije Universiteit Brussel \\
    	School of Business and Economics, Vrije Universiteit Amsterdam 
    	\and
    	S\'{e}bastien Laurent \\
    	% Aix-Marseille University (Aix-Marseille School of Economics) \\ 
    	% CNRS \& EHESS \\
    	% Aix-Marseille Graduate School of Management-IAE 
    	Aix Marseille Univ., CNRS, AMSE, Marseille, France\\
    	Aix-Marseille Graduate School of Management-IAE 
    	\and
    	Christopher J. Neely \\
    	Research Division, Federal Reserve Bank of St. Louis
    	}
	
    \date{\normalsize{\today}}
	\maketitle

%%%%%%%%%%%%%%%%%%%%%%%%%%%%%%%%%%%%%%%%%%%%%%%%%%%%%%%%%%%%%%%%%%%%%%%%%%%%%%

\vspace{-21pt}
\begin{abstract}
	Stock prices often react sluggishly to news, producing gradual jumps and jump delays. 
    \if1\comment
    \blue{Something something noise.}
    % See Word file with Chris. 
    % If trades do not occur at the time a jump in the underlying efficient price occurs, then observed news reactions can be sluggish because trading is not continuous even if market participants are constantly aware of fundamentals.
    % price discovery?
    \blue{Noise ... Changing time labels.} 
    \fi
    % Market participants must trade to 
    % reveal private information and 
    % reach a consensus about he impact of some piece of news. 
    Econometricians typically treat these sluggish reactions as microstructure effects and settle for a coarse sampling grid to guard against them. 
    % the frictions with which actual trades take place.  
    % A coarse sampling grid guards against these microstructure effects, the frictions with which actual trades take place, but oversmooths actual changes. 
    % 
	Synchronizing mistimed stock returns % of the Dow 30 
    on a fine sampling grid 
    allows us to 
    automatically detect noisy jumps and 
    better approximate 
    % recover 
    the true common jumps in related stock prices. 
    \iffalse
	and improves out-of-sample % covariance forecasts and 
    portfolio performance. 
    \fi
    \if1\comment
    % CN: 
	\blue{And we find what earth-shaking result? (We document jump noise, ...)}	
    % Try different things. 
    % FOMC statements. 
    \fi
    % We document that 
    % During some difficult-to-interpret FOMC statements, 
    % some Dow 30 stock prices jump simultaneously when news is released and for some stocks a sequence of transactions reveal news in the price. 
    % it took Dow 30 stocks up to four minutes of transactions to reveal the news in the price. 
\end{abstract}

\noindent%
    
%    \vfill

%%%%%%%%%%%%%%%%%%%%%%%%%%%%%%%%%%%%%%%%%%%%%%%%%%%%%%%%%%%%%%%%%%%%%%%%%%%%%%
% \spacingset{1.45}
%%%%%%%%%%%%%%%%%%%%%%%%%%%%%%%%%%%%%%%%%%%%%%%%%%%%%%%%%%%%%%%%%%%%%%%%%%%%%%

\section{Introduction}\label{secIntro}

Major economic news, such as pre-scheduled announcements, natural disasters or geopolitical conflicts, trigger common jumps in related stock prices \citep[see \textit{e.g.}][for some empirical examples]{li2017mixed}. 
Statistical tests for these common jumps, or so-called  ``cojump tests", implicitly assume that jumps occur simultaneously in relevant assets but, in fact, jumps occur asynchronously in transaction prices.
Stock prices can move sluggishly  \citep{bandi2017excess}, jumps can be gradual \citep{barndorff2009realized} and jumps of less-liquid individual assets typically lag those of the more-liquid market index \citep{li2017mixed}. 
Most researchers have dealt with this problem by settling for a coarse sampling grid \citep[see \textit{e.g.},][]{barndorff2009realized,bollerslev2008risk,lahaye2011jumps,li2019rank}. 
% Barndorf-Nielsen: add jumps together. 
Such a coarse grid guards against microstructure effects, the frictions with which actual trades take place, but it is restrictive in that it oversmooths actual changes \citep[see \textit{e.g.},][]{ait2004disentangling}. 
% Coarse grids are unsatisfactory, because, the finer the sampling grid, the higher the probability that a jump can be recognized as such \citep{ait2004disentangling}. 

\iffalse
\blue{\textbf{SL:} Add JE paper on cojump networks.}
\fi

\iffalse
\if1\comment
\blue{Add a paragraph on this type of noise to connect our paper to the broader ``market microstructure noise" literature.}
% See Tex\introduction
\blue{Microstructure noise may exhibit rich dynamics depending on its source. 
Sluggishness in news reactions is another type of noise. 
}
\fi
% Microstructure noise contaminates
% Prices can also be sluggish. 
Frictions in financial markets may cause observed prices to deviate from the underlying underlying equilibrium (often called ``efficient") price. 
Features such as tick size, discrete observations, bid-ask spreads, adverse selection, liquidity and inventory control produce market microstructure noise 
% and contaminate the underlying (often called ``efficient") price 
\citep[see \textit{e.g.},][]{christensen2014fact, leemykland2012jumps}. 
Prices may also be sluggish because market participants must trade to reveal private information and reach a consensus about the impact of some piece of news. 
% If trades do not occur at the time a jump in the underlying equilibrium 
% (often called ``efficient") 
% price occurs, then observed news reactions appear to be sluggish because trading is not continuous even if market participants are constantly aware of fundamentals.
\fi

We offer an alternative strategy which changes the time labels of some financial time series observations on a fine sampling grid to approximately recover the efficient 
common jump in a basket of stocks. 
% using % only 
% noisy transaction prices. % on a fine sampling grid. 
Asynchronous impoundment of news causes the value of a synthetic stock index to deviate from the price of an exchange-traded fund (ETF),  even though they consist of the same stocks. 
Assuming that an ETF price  tracks the latent, equilibrium 
% (often called)
(often called ``efficient") 
value of a stock index, the spread between the value of a synthetic index and the ETF price measures 
%  the collective misalignment of the noisy stock prices with their respective equilibrium levels. 
the sluggishness in news reactions. 
% We minimize the spread by synchronizing the jump in the ETF with the corresponding cojump in the underlying stocks. 
Combinatoric methods rearrange jumps to minimize the spread and approximately recover the latent efficient price. 

To rearrange stock jumps, we extend the pioneering work of \citet{puccetti2012computation} and \citet{embrechts2013model} on rearrangements. 
Their rearrangement algorithm is best known as an actuarial tool to bound portfolio risk, but it can also be applied to
other disciplines, such as operations research \citep[see \textit{e.g.},][]{boudt2018block}. 
Rearrangements can also synchronize stock jumps and recover the common jump on a fine sampling grid, provided we penalize economically implausible rearrangements.  % using a mixed-integer linear program. 
For example, we only allow for a rearrangement of jumps backward in time because we assume that stock prices are sluggish and lag the highly liquid and carefully watched ETF, they do not lead it.

We apply our methods to investigate the reactions of Dow 30 stock prices in event windows around DIA ETF jumps. 
For example, the Federal Reserve announced rate cuts on September 18, 2007,  at 14:15 US Eastern Time, after which markets % took time 
took up to five minutes % 4+1
to incorporate the Fed’s news into the Dow 30 stock’s prices. 
Rearrangements synchronize 19 (out of 23) scattered stock jumps with the ETF jump, approximately recovering the common jump in the stocks. This is not a stand-alone event: the rearrangement linear program rearranges stock jumps in 180 cases. 

\if1\comment
\blue{And we find what earth-shaking result? Add summary statistics and a convincing application.}
% Something on sampling observations? 
% Iets over de schatters... 
% Other estimators, like the multivariate realized kernel in Barndorff-Nielsen et al. (2011) or a Cholesky factorization in Boudt et al. (2017), protect against mild market microstructure noise and the Epps (1979) effect, that is, the downward bias in covariance estimates due to asynchronous trading. Using rearranged returns in (32) also protects against asynchronous jumps and the underestimation of jump dependence.
\fi
Synchronizing mistimed stock returns 
% also 
improves estimates of the daily realized covariance matrix. 
Other estimators, like the multivariate realized kernel in 
\citet[][]{barndorff2011multivariate} or a Cholesky factorization in \citet{boudt2017positive},  protect against  mild market microstructure noise 
and the \citet{epps1979comovements} effect, that is, the downward bias in 
covariance estimates due to asynchronous trading. 
% Using rearranged returns % in \eqref{eqRC} 
But rearranging returns 
protects against
% asynchronous jumps and 
the underestimation of jump dependence due to asynchronous jumps 
and 
% Rearrangements  
improves the 
out-of-sample financial performance compared to using raw returns. 
% Not convincing enough.

We proceed as follows. 
Section \ref{secFramework} details the synchronization method using a toy example. 
Section  \ref{secEmpirics} illustrates an empirical example of a rearranged sluggish cojump in the Dow 30 and includes a portfolio allocation exercise. 
Section \ref{secConclusion} concludes. 

%%%%%%%%%%%%%%%%%%%%%%%%%%%%%%%%%%%%%%%%%%%%%%%%%%%%%%%%%%%%%%%%%%%%%%%%%%%%%%%%%%%%%%%%%%%%%%%%%%%%%%%%%%%%%%%%%%%%%%%%%

\section{Synchronizing jumps: A combinatorial problem}
\label{secFramework}

% Barndorff-Nielsen start met Non-synchronous trading ... 
A salient feature of multivariate high-frequency financial data is the occurence of non-synchronous trading; 
it is rare for any two assets to trade simulataneously. 
% Any two assets rarely trade at the same instant. 
% Non-synchronous trading delivers 
This leads to 
prices at irregularly spaced times, differing across assets. 
Addressing asynchronicity through the coordinated collection of multivariate data 
has been an active area of research in financial econometrics in recent years, see \textit{e.g.}, 
 \citet{barndorff2011multivariate} or \citet{boudt2017positive} 
and the references therein, and the concept of so-called ``stale" prices has been integral to covariance estimations since  \citet{epps1979comovements}. 
Nonetheless, the state-of-the-art sampling schemes like refresh-time sampling \citep{barndorff2011multivariate},  are not tailored to price jumps, with asynchronous jumps not necessarily resulting from non-synchronous trading. 
At times, prices may be ``sluggish"; the asset might be trading, but due to various factors, 
the news might not yet be impounded in the price. 
To address this problem, we synchronize the timing of multivariate jumps using what we call ``Jump Sampling". 
% Refresh-time samplign is not tailored to jumps.
% Asynchronous jumps are not due to non-synchronous trading. In some cases, the asset is trading, but the news has not yet been impounded in the price, for various reasons. 
% Synchronize the timing of multivariate jumps using what we call Jump Sampling. 
% Sampling scheme.  
This technique refines the detection of high-frequency cojumps and, in turn,  the realized covariance matrix. 
% coordinat ethe collection of data using refresh-time. 
% staleness in the data. 

% Stock prices often react sluggishly to news, producing gradual jumps and jump delays. That is, a jump in the underlying equilibrium price may not be immediately reflected in the observed price due to various trading frictions. 

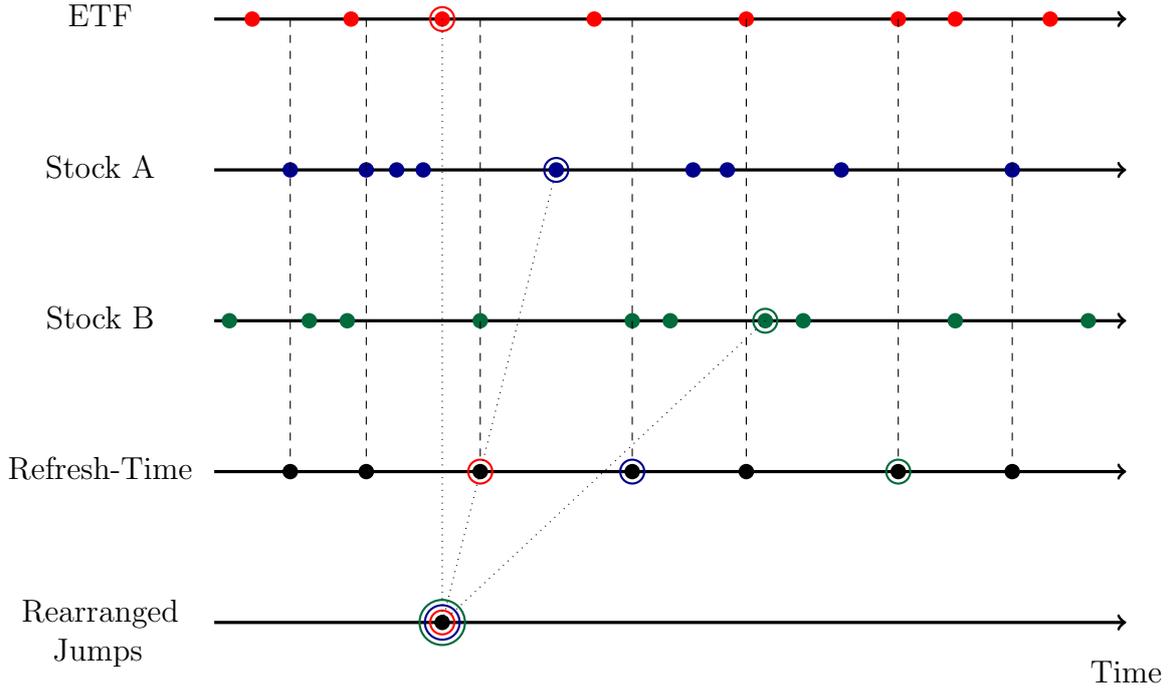
\begin{figure}[htb!]
	\centering
	\caption{Jump sampling}
	\label{figJumpSample}
	\scalebox{1.}{
		\vspace{.3cm}
		\begin{tikzpicture}[xscale=1.]
			
			%% Asset 1
			\node[align=left] at (-4.5, -.1) {ETF};
			\draw [very thick,->] (-3,0) -- (9,0);
			%% circles
			\filldraw [red] (-2.5,0) circle (2.5pt);
			\filldraw [red] (-1.2,0) circle (2.5pt);
			\filldraw [red] (0,0) circle (2.5pt);
			\draw [red] (0,0) circle (4.5pt); % JUMP
			\filldraw [red] (2,0) circle (2.5pt);
			\filldraw [red] (4,0) circle (2.5pt);
			\filldraw [red] (6,0) circle (2.5pt);
			\filldraw [red] (6.75,0) circle (2.5pt);
			\filldraw [red] (8,0) circle (2.5pt);
			
			%% Asset 2
			\node[align=left] at (-4.5, -2.1) {Stock A};
			\draw [very thick,->] (-3,-2) -- (9,-2);
			\filldraw [darkblue] (-2,-2) circle (2.5pt);
			\filldraw [darkblue] (-1,-2) circle (2.5pt);
			\filldraw [darkblue] (-.6,-2) circle (2.5pt);
			\filldraw [darkblue] (-.25,-2) circle (2.5pt);
			\filldraw [darkblue] (1.5,-2) circle (2.5pt);
			\draw [darkblue] (1.5,-2) circle (4.5pt); % JUMP
			\filldraw [darkblue] (3.3,-2) circle (2.5pt);
			\filldraw [darkblue] (3.75,-2) circle (2.5pt);
			\filldraw [darkblue] (5.25,-2) circle (2.5pt);
			\filldraw [darkblue] (7.5,-2) circle (2.5pt);
			
			%% Asset 3
			\node[align=left] at (-4.5, -4.1) {Stock B};
			\draw [very thick,->] (-3,-4) -- (9,-4);
			\filldraw [cadmiumgreen] (-2.8,-4) circle (2.5pt);
			\filldraw [cadmiumgreen] (-1.25,-4) circle (2.5pt);
			\filldraw [cadmiumgreen] (-1.75,-4) circle (2.5pt);
			\filldraw [cadmiumgreen] (.5,-4) circle (2.5pt);
			\filldraw [cadmiumgreen] (2.5,-4) circle (2.5pt);
			\filldraw [cadmiumgreen] (3,-4) circle (2.5pt);
			\filldraw [cadmiumgreen] (4.25,-4) circle (2.5pt);
			\draw [cadmiumgreen] (4.25,-4) circle (4.5pt); % JUMP
			\filldraw [cadmiumgreen] (4.75,-4) circle (2.5pt);
			\filldraw [cadmiumgreen] (6.75,-4) circle (2.5pt);
			\filldraw [cadmiumgreen] (8.5,-4) circle (2.5pt);
			%	
			%% Refresh-Time
			\node[align=left] at (-4.5, -6.1) {
				% \, \, Standard \\  
				Refresh-Time};
			\draw [very thick,->] (-3,-6) -- (9,-6);
			% black.. We don't care about any asset. 
			% tau1
			\filldraw [black] (-2,-6) circle (2.5pt);
			\draw [thin,dashed] (-2,0) -- (-2,-6);
			% tau2
			\filldraw [black] (-1,-6) circle (2.5pt);
			\draw [thin,dashed] (-1,0) -- (-1,-6);
			%tau3
			\filldraw [black] (.5,-6) circle (2.5pt);
			\draw [red] (.5,-6) circle (4.5pt); % Jump 
			\draw [thin,dashed] (.5,0) -- (.5,-6);
			%tau4
			\filldraw [black] (2.5,-6) circle (2.5pt);
			\draw [darkblue] (2.5,-6) circle (4.5pt); % JUMP
			\draw [thin,dashed] (2.5,0) -- (2.5,-6);
			%tau5
			\filldraw [black] (4,-6) circle (2.5pt);
			\draw [thin,dashed] (4,0) -- (4,-6);
			%tau6
			\filldraw [black] (6,-6) circle (2.5pt);
			\draw [cadmiumgreen] (6,-6) circle (4.5pt); % Jump
			\draw [thin,dashed] (6,0) -- (6,-6);
			%tau7
			\filldraw [black] (7.5,-6) circle (2.5pt);
			\draw [thin,dashed] (7.5,0) -- (7.5,-6);

			\node[align=left] at (-4.5, -8.5) {Rearranged \\
				\, \, Jumps};
			\draw [very thick,->] (-3,-8) -- (9,-8);
			% colour, we sample univariately
			% ASSET 1: 
			\draw [red] (0,-8) circle (4.5pt); % JUMP
			\filldraw [black] (0,-8) circle (2.5pt); % JUMP
			% ASSET 2:
			\draw [darkblue] (0,-8) circle (6.5pt); % JUMP
			%	% Asset 3
			\draw [cadmiumgreen] (0,-8) circle (8.5pt); % JUMP
			% \filldraw [cadmiumgreen] (4.25,-8) circle (2.5pt); % JUMP
			%	\draw [semithick,dotted] (4.25,-4) -- (4.25,-8);
			\draw [thin,dotted] (0,0) -- (0,-8); % ETF
			
			% CN add lines. 
			\draw [thin,dotted] (1.5,-2) -- (0,-8); % Stock A
			\draw [thin,dotted] (4.25,-4) -- (0,-8); % Stock B 4.25,-4
			
			\node[align=right] at (9.0, -8.8) {Time};
			
		\end{tikzpicture}
	}
	
	 			\vspace{.3cm}
				\begin{minipage}{1.0\linewidth}
					\begin{tablenotes}
						\small
						\item {
							\medskip
							Note: 
							% Non-synchronous trading delivers prices at irregularly spaced times, differing across assets. 
							This figure compares refresh-time sampling to  jump sampling  in the presence of asynchronous observations and jumps. 
							%
							% This figure is inspired by the well-known Figure 1 in \citet*[][]{barndorff2011multivariate} in which the authors illustrate refresh-time in a situation with three assets (without the occurence of jumps). 						
							In each asset's case, the filled dots indicate when the posted prices were updated, and the open dot represents the time at which the price jumps. % of three particular assets -- 
							The vertical dashed lines represent the sampling times generated from the three assets using  refresh-time sampling.
							%
							% We expand upon this concept to include scenarios with price jumps,
							% what happens using refresh-time sampling in the presence of asynchronous jumps in the case of three particular assets: 
							% a basket instrument and its two components -- jump. 
							% For each of the assets, the filled dots represent the times at which the posted prices have been updated. 
							% 
							% The open dots correspond to a jump arrival in each of the assets. 
							%
							% The vertical dashed lines represent the sampling times generated from the three assets with refresh-time sampling. 
							% For example, the first black dot represents the time it has taken for all the assets to trade. 
							%
							% Refresh-time with jumps correspond to sampling the jumps when all the assets have jumped. 
							% which introduces by construction a cojump. 
							% Even worse, in some cases, not all assets will jump. 
							% 
							% Our local rearrangement sampling approach links the jump of the index with the underlying stocks. 
							Jump sampling rearranges mistimed jumps to occur simultaneously with the ETF jump. 
							% The vertical dotted lines correspond to our univariate jump sampling approach, which \textit{presedes} the local rearrangement of jumps: we univariately sample the jump arrival in each of the assets. 
							% 
							%				\red{Refresh-time with jumps erbij zetten; je gaat altijd cojumps hebben...  Inleiding, is wat raar omdat het raar staat... Je kan het ook gebruiken voor jump delayy te illustreren in uw schatting. Springt er wat tussen uit. Benchmark.. Ik wil schatten of mijn asset te laat is of niet. Hoort niet bij de cojump... puur gefocust op de jumps... Dat is onze meerwaarde...}
							%				\textbf{\red{Figuur naar de empirie... }}
							%				\red{Kris ziet de relevantie van Refresh-Time Samplign niet}
							%				% Het is wel duidelijk. Hier zien we niet welke jumps te maken hebbe,n met de index jump. Bij de onze wel... 
							%				% Duidelijke link naar BNS.  
							%				%% Multiple testing paper Goeman: It has the potential to ... 
							% \red{Klopt niet helemaal. We detecteren jumps niet op de hoogtste frequencie. Maar op minuten... }
						}
						%% KB
						% \red{figuur nog relevant? met mispricing en sluggish jumps lijkt de figuur niet meer ok}
						% Dat denk ik wel; heb de laatste rij weer toegevoegd. It's about the sampling which presedes the local rearrangement. Laatste rij heb ' k weggelaten omdat dat veel informatie was. Staat er nu weer in. 
						% 
						% }
					\end{tablenotes}
				\end{minipage}
\end{figure}

Figure \ref{figJumpSample} compares refresh-time sampling to  jump sampling  in the presence of asynchronous observations and jumps. 
It draws inspiration from the well-known Figure 1 in \citet[][]{barndorff2011multivariate}, illustrating refresh-time in a situation with three assets (without the occurence of jumps). 		
We expand upon this concept to include scenarios with asynchronous price jumps, focusing on three specific assets:  a basket instrument and its two underlying stocks. 
In each asset's case, the filled dots indicate the updates in posted prices, and an open dot pinpoints the time at which the price jumps. 
% what happens using refresh-time sampling in the presence of asynchronous jumps in the case of three particular assets: 
%  -- jump. 
Vertical dashed lines represent the sampling times generated from the three assets, using the refresh-time sampling approach. For example, the first black dot represents the time it has taken for all three assets to trade. 
% Our approach differs from refresh-time sampling % in important ways. 
% in an important way. 
%
%
% For each of the assets, the filled dots represent the times at which the posted prices have been updated. 
% 
% The open dots correspond to a jump arrival in each of the assets. 
%
% The vertical dashed lines represent the sampling times generated from the three assets with refresh-time sampling. 
% For example, the first black dot represents the time it has taken for all the assets to trade. 
%
% Refresh-time with jumps correspond to sampling the jumps when all the assets have jumped. 
% which introduces by construction a cojump. 
% Even worse, in some cases, not all assets will jump. 
But because asynchronous jumps are not due to (il)liquidity issues, refresh-time sampling does not resolve the asynchronicity inherent in the jumps. 
% 
% Our local rearrangement sampling approach links the jump of the index with the underlying stocks. 
As a solution, we introduce a new jump sampling scheme, which rearranges mistimed jumps to occur simultaneously with the ETF jump. 

% \clearpage

In what follows, we detail how we synchronize stock jumps using combinatorics.  
We optimally rearrange jumps, penalizing economically implausible rearrangements. 
A simulated example clarifies the mechanics of the rearrangements. 

\subsection{A DGP for sluggish news reactions}

We assume a data generating process for the sluggish prices of the stocks in the index, which features gradual jumps and jump delays.\footnote{Gradual jumps are when the prices exhibit strong linear trends for periods of a few minutes \citep{barndorff2009realized}. Jump delays are when jumps of individual assets follow those of the highly liquid market index during market-wide events \citep{li2017mixed}.}  
A jump in the underlying equilibrium price may not be immediately reflected in the observed price due to various trading frictions. Such complications are not captured in the standard martingale-plus-noise price model, but they are important in the empirical analysis of multivariate jump processes.

Let ${X}_t = (X_{1,t}, ..., X_{p,t})^\top$ denote the logarithmic $p$-variate, equilibrium  (or so-called ``efficient") price of the $p$ stocks in the market index. The price process is defined on a filtered probability space $(\Omega, \mathcal{F}, (\mathcal{F})_{t \geq 0}, \mathbb{P})$ and is adapted to the filtration $\mathcal{F}_t$ that represents information available to market participants at time $t$, with $t \geq 0$. We assume that ${X}$ operates in an arbitrage-free, frictionless market, which implies that ${X}$ is a semimartingale. Econometricians \citep[see][]{ait2014high} model stock prices ${X}$ as a jump-diffusion process, which includes a continuous Brownian component and a discontinuous jump component: 
\begin{equation}
	\begin{aligned}
		\label{eqEfficientPrice}
		{X}_t 
		&= {X}_t^c + {X}_t^d, 
		\, \text{with}, \, \, 
		\\
		{X}_t^c &\equiv 
		{X}_0 
		+ 
		\int_0^t {b}_s ds
		+ 
		\int_0^t {\sigma}_s d{W}_s,  
		\\
		{X}_t^d &\equiv 
		\textstyle \sum_{s \leq t} \Delta {X}_s, 
	\end{aligned}
\end{equation} 
in which 
$t \geq 0$, ${b}$ is the drift process, ${\sigma}$ is the stochastic (co)volatility process, ${W}$ is a multivariate Brownian motion and $\Delta {X}_t \equiv {X}_t - {X}_{t-}$, with ${X}_{t-}$ the left limit at time $t$,  denotes the jumps of ${X}$ at time $t$. 
A stock's growth prospects generates a jump in single stock price. 
Major economic news, such as pre-scheduled announcements, natural disasters or geopolitical conflicts, trigger  common (\textit{i.e.} synchronous) jumps in related stock prices \citep[see \textit{e.g.}][for some empirical examples]{li2017mixed}.

In practice we do not observe the price process in \eqref{eqEfficientPrice}. 
Instead we observe discretely sampled, noisy transaction prices. 
Frictions such as tick size, discrete observations, bid-ask spreads, adverse selection, liquidity and inventory control produce market microstructure noise \citep[see \textit{e.g.},][]{christensen2014fact, leemykland2012jumps,li2022remedi}.
Prices may also be sluggish because market participants must trade to reveal private information and reach a consensus about the impact of some piece of news. 
If trades do not occur at the time a jump in the underlying efficient price occurs, then observed news reactions can be sluggish because trading is not continuous even if market participants are constantly aware of fundamentals.

We model the observed log price process ${Y}_t = (Y_{1,t}, ..., Y_{p,t})^\top$ of the $p$ stocks as 
% the sum of a contaminated Brownian component and a contaminated jump component: 
contaminated version of \eqref{eqEfficientPrice} observed at discrete intervals: 
\begin{equation}
	\begin{aligned}
		\label{eqObservedPrice}
		{Y}_{i\Delta_n}  
		&= {Y}^c_{i\Delta_n} + {Y}^d_{i\Delta_n}, 
		\, \text{with}, \, \, 
		\\
		{Y}_{i\Delta_n}^c 
		&\equiv {X}^c_{i\Delta_n} + {u}_{i\Delta_n}, 
		\\
		{Y}_{i\Delta_n}^d 
		& \equiv 	
		\textstyle \sum_{h\Delta_n \leq i\Delta_n} \Delta {Y}_{h\Delta_n}.
	\end{aligned}
\end{equation}
There are two kinds of noise: microstructure noise and mistimed jumps. 
Microstructure noise ${u}$ contaminates the efficient price process ${X}$, but is typically too small to substantially contaminate the discontinuous part ${X}^d$. It can neither generate gradual jumps \citep[as in][]{barndorff2009realized} nor jump delays \citep[as in][]{li2017mixed}. 
% In practice, econometricians view gradual jumps and jump delays as a microstructure-type effect, related to the difference in liquidity in the trading of different types of assets \citep{li2019rank}. 
We capture the mistimed or mismeasured jumps in a separate noisy jump component ${Y}^d$, which allows a sluggish news reaction, spreading the  stock
jump across several time intervals. 

\begin{figure}[htb!]
	\caption{The discontinuous component of a stock's price may react sluggishly to news}
	\centering
	
	\includegraphics[width=.90\textwidth,angle = -0]{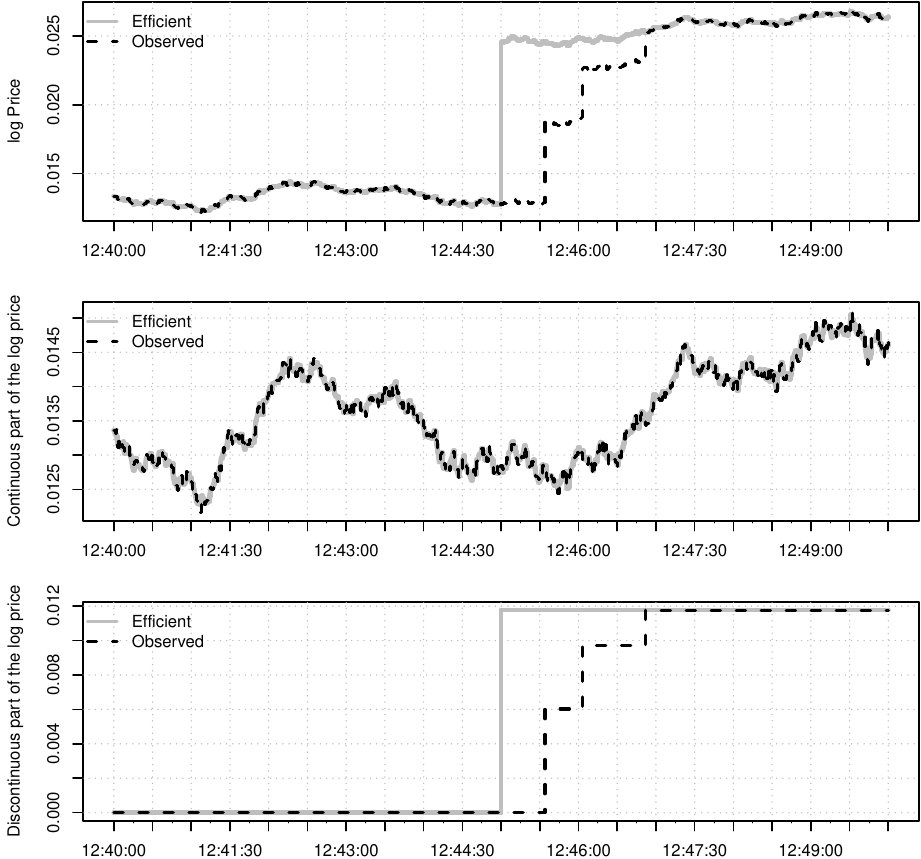}
	\label{fig:Simulate-One-Path}
	\begin{minipage}{1.0\linewidth}
		\begin{tablenotes}
			\small
			\item {
				\medskip
				Note: 
				We plot the decomposition of the efficient \eqref{eqEfficientPrice} and observed \eqref{eqObservedPrice} log price process for one sample path 				during a short event window. 
				We simulate second-by-second ($\Delta_n = 1/23,401$) efficient and observed prices for 1 stock across one trading day.  
				The top panel shows the log price. The middle panel shows the continuous part of the log price, and the bottom panel shows the discontinuous part of the log price. 
				The efficient stock price jumps at $i\Delta_n = 11,701$ or
				12:45:00.
				Mild market microstructure noise contaminates the continuous part of the price process, while delays in the jump process contaminate the discontinuous part of the price process.
			}
		\end{tablenotes}
	\end{minipage}
\end{figure}

Figure \ref{fig:Simulate-One-Path} shows a simulated sample path of this new DGP for 1 stock. 
The top panel of Figure \ref{fig:Simulate-One-Path} illustrates that the efficient stock price jumps at 12:45 in reaction to news. 
In the following 112 seconds, the observed price (in black) catches up with the new equilibrium level (in gray) by gradually matching the jump. 
The middle and bottom panels respectively decompose the price process into its continuous Brownian component and its jump component. 
The middle panel compares the efficient, continuous price with the one contaminated by mild market microstructure noise. 
The bottom panel compares the efficient, sudden jump with the contaminated, gradual jump. 
Our assumed DGP uses a step function to model how observed prices incorporate news.
Appendix \ref{secSimExample} shows how to spread the jump across several time intervals. 

The observed price process \eqref{eqObservedPrice} combines the frictions and the sampling frequency. 
We sample discretely at time points $i\Delta_n$, with $i = 0, ..., \lfloor T / \Delta_n \rfloor$, across a time span $T$, in which $\lfloor \cdot \rfloor$ denotes the floor function. 
%  In practice, econometricians view gradual jumps and jump delays as a microstructure-type effect, related to the difference in liquidity in the trading of different types of assets \citep{li2019rank}. To guard against such microstructure effects, econometricians sample returns on a coarse sampling grid or manually replace the gradual jump by a sudden one. 
There is less noise on a coarse sampling grid, \textit{i.e.} at a lower sampling frequency $\Delta_n$, but data at such lower frequencies tend to oversmooth actual changes \citep[as in][]{bollerslev2008risk, lahaye2011jumps, li2019rank}. 
% Synchronizing scattered jumps on a fine sampling grid allows us to automatically detect noisy jumps and better approximate the true common stock jump component. 
The finer the sampling grid, the higher the probability that a jump can be recognized as such \citep{ait2004disentangling}. 

\subsection{Collecting asynchronous jumps in a jump-event matrix}
\label{ssecJumpMatrix}

% We investigate the relationship between a jump in an index ETF, \textit{i.e.} an index tracker, and the corresponding cojumps in the underlying basket of stocks on a fine sampling grid. 
When multiple stocks react sluggishly to new information, their jumps are 
asynchronous % across stocks 
on a fine sampling grid, 
and these jumps will generally not coincide with the jump in the price of an index tracker.
%  on a fine sampling grid. % ETF. 
Empirical evidence corroborates this prediction: jumps of less-liquid individual assets typically lag those of the more-liquid market index
\citep{li2017mixed} and the ETF jumps more often than a synthetically constructed index of stocks \citep{bollerslev2008risk}. 
% We introduce a new object, the jump-event matrix, to analyze the asynchronous jumps in the index ETF 
% , i.e., an index tracker, 
% and the corresponding jumps in the underlying basket of stocks.

\subsubsection{The spread measures sluggishness in high-frequency data}
\label{secMispricing}

Let $w_{k,t}$, with $k = 1, ..., p$, be the weights allocated to each stock in the market index at each moment in time. 
The price of the synthetically constructed  index portfolio, $S_{i\Delta_n}$, is a linear combination of the observed stock prices in (\ref{eqObservedPrice}), sluggishly incorporating its jump component: 
\begin{align}
	\label{eqSynPrice}
	S_{i\Delta_n} &= \sum_{k = 1}^p {w}_{k,i\Delta_n} Y_{k,i\Delta_n}.
\end{align}

An ETF log price process, $Z_t$, tracks an index of the $p$ stocks. 
We assume that the observed log price $Z$ replicates a portfolio of efficiently priced stocks \eqref{eqEfficientPrice}, efficiently incorporating its jump component%
\footnote{To simplify our notation, we rely on a weighted average of individual log returns as opposed to simple returns. This difference is considered as minor in empirical applications \citep*[][pp. 9]{jondeau2007financial}.}:
\begin{align}
	\label{eqETFPrice}
	Z_{t} &= \sum_{k = 1}^p {w}_{k,t} X_{k,t}.
\end{align}

The deviation or 
spread in prices is the difference between the observed prices on a synthetic index of stocks \eqref{eqETFPrice} and the prices of an observable ETF trading the index \eqref{eqSynPrice}: 
\begin{align*}
	% \label{eqPriceSpread}
	\delta^p_{i\Delta_n} 
	&:=
	S_{i\Delta_n} - Z_{i\Delta_n}. 
\end{align*}
Similarly, we can also define the spread in returns as the difference between the observed returns on a synthetic index of stocks $\Delta^n_i S := \sum_{k = 1}^p {w}_{k,i\Delta_n} \Delta^n_i {Y}_k$ 
and the returns of an observable ETF trading the index $\Delta^n_i Z := {Z}_{i \Delta_n} - {Z}_{(i-1)\Delta_n}$ or, equivalently, the percentage change of the price spread in (\ref{eqPriceSpread}):
\begin{align}
	\label{eqPriceSpread}
	\delta^r_{i\Delta_n}
	:=
	\delta^p_{i\Delta_n} - \delta^p_{(i-1)\Delta_n}
	= 
	\Delta^n_i S  - \Delta^n_i Z.
 \end{align}

We expect the ETF log price $Z$ to nearly equal the synthetic log price $S$ in the absence of sluggish prices. 
Only microstructure noise would separate the two prices. However, asynchronous jumps cause the price of the synthetic index portfolio to deviate from the presumably efficient price of an ETF tracking the index.\footnote{Within our theoretical model descriptions, the difference between the price of the synthetic index and the ETF price \eqref{eqPriceSpread} equals microstructure noise component minus the discontinuous component that has not yet been impounded into the observed prices: 
\begin{align}
	\label{eqPriceSpreadTheory}
	S_{i\Delta_n} - Z_{i\Delta_n}
	&= 
	\sum_{k = 1}^p {w}_{k,i\Delta_n} {u}_{k,i\Delta_n} 
	+ \sum_{k = 1}^p {w}_{k,i\Delta_n} 
	\left( 
	{Y}^d_{k,i\Delta_n} - {X}^d_{k,i\Delta_n}
	\right).
\end{align}
	This sharp theoretical decomposition is unobservable to the econometrician in empirical data. 
}
The sluggish components of jumps are much larger than microstructure noise, so asynchronous impoundment of news % dominates the large price spreads.
drives the spread in prices. 
Hence, the spread \eqref{eqPriceSpread} between the ETF price and a synthetically constructed index measures the collective misalignment  
of noisy stock prices with their % respective 
efficient levels. 
The goal is to rearrange jumps in empirical data to minimise the spread and recover the latent efficient price. 

To illustrate the workings of our procedures, we consider a stylistic 3-stock universe $(p = 3)$ and a corresponding ETF, in which stock prices vary in how quickly they impound news. The stock names A, B and C correspond to the indices $k = 1, 2$ and $3$. The ABC ETF price is an equally weighted average of the  underlying  stocks' efficient prices \eqref{eqETFPrice} and the synthetic ABC portfolio is an equally weighted average of the stocks' observed prices \eqref{eqSynPrice}. 
The sampling frequency is one minute ($\Delta_n = 1/391$).

Consider  the following time series of return vectors of the three stocks and the ABC ETF: 
\begin{align}
	\label{eqExampleReturns}
	\Delta^n_i {Y}_1 = 
	\begin{bmatrix*}[c]
		\phantom{-}\vdots \\
		-0.018 \\ -0.031 \\ -0.057 \\ \phantom{-}\underline{0.629} \\ \phantom{-}\underline{0.651} \\
		\phantom{-}\vdots
	\end{bmatrix*}, \, 
	\Delta^n_i {Y}_2 = 
	\begin{bmatrix*}[c]
		\phantom{-}\vdots \\
		\phantom{-}0.015 \\ -0.067 \\ -0.029 \\ \phantom{-}\underline{1.201} \\ \phantom{-}0.062 \\
		\phantom{-}\vdots
	\end{bmatrix*}, \, 
	\Delta^n_i {Y}_3 = 
	\begin{bmatrix*}[c]
		\phantom{-}\vdots \\
		-0.120 \\ -0.104 \\ \phantom{-}0.088 \\ \phantom{-}0.017 \\ \phantom{-}0.074 \\
		\phantom{-}\vdots
	\end{bmatrix*}, \, \text{and} \, 
	\Delta^n_i {Z} = 
	\begin{bmatrix*}[c]
		\phantom{-}\vdots \\
		-0.039 \\ -0.071 \\ \phantom{-}\underline{0.807} \\ \phantom{-}0.001 \\ \phantom{-}0.073 \\ 
		\phantom{-}\vdots
	\end{bmatrix*},
\end{align}
in which the returns are reported in percentages and the jump returns are underlined. 
Prices asynchronously incorporate news. 
The first column shows that Stock A jumps gradually and finishes its jump 2 minutes after the ETF in the last column, 
while the second column shows that stock B's jump is not gradual but 1 minute late and stock C does not jump (third column). 

These delays cause the implied (inefficient) returns of the ABC portfolio
to deviate from the efficient ABC ETF returns. 
The sum of the first three columns of the matrix below, which calculates the return spread $\delta_{i\Delta_n}^r$, is the equally weighted, linear combination of the log returns of stocks A, B, and C (the first three columns in \eqref{eqExampleReturns}), that is, the return on the synthetic ABC portfolio. 
The fourth column on the left side of the equal sign is the log return of the ABC ETF (the last column in \eqref{eqExampleReturns}).
Their difference, on the right-hand side of equation \eqref{eqReturnSpreadExample}, is the return spread \eqref{eqPriceSpread}, in each of the five periods.
\begin{align}
	\label{eqReturnSpreadExample}
	\delta^r_{i\Delta_n} 
	= 
	\left[
	\begin{matrix*}[c]
		\phantom{\vdots} \\
		\phantom{.} \\ % 1
		\phantom{.} \\ % 2
		\phantom{.} \\ % 3
		\phantom{.} \\ % 4
		\phantom{.} \\ % 5
		\phantom{\vdots} \\
	\end{matrix*}
	\right.
	\underbrace{
		\begin{matrix*}[r]
			\phantom{\vdots} \\
			\nicefrac{1}{3}.   (         {-}0.018   + 0.015   - 0.120) \\
			\nicefrac{1}{3}.   (         {-}0.031   - 0.067   - 0.104)    \\
			\nicefrac{1}{3}.   (         {-}0.057   - 0.029   + 0.088)  \\
			\nicefrac{1}{3}.   (\phantom{{-}}\underline{0.629}  + \underline{1.201}   + 0.017)  \\
			\nicefrac{1}{3}.   (\phantom{{-}}\underline{0.651}  +  0.062  + 0.074)  \\
			\phantom{\vdots} \\
		\end{matrix*}
	}_{\substack{\text{{ABC portfolio}} \\ \text{{returns}} }}
	% white space. the minuses are too close
	\begin{matrix*}[c]
		\phantom{\vdots} \\
		\phantom{.} \\
		\phantom{.} \\
		\phantom{.} \\
		\phantom{.} \\
		\phantom{.} \\
		\phantom{\vdots} \\
	\end{matrix*}
		\begin{matrix*}[c]
		\vdots \\
		{-} \\ % 1 
		{-} \\ % 2
		{-} \\ % 3
		{-} \\ % 4
		{-} \\ % 5
		\vdots
		\end{matrix*}
	\underbrace{
		\begin{matrix*}[c]
			\phantom{\vdots} \\
			({-}0.039) \\
	        ({-}0.071) \\
			\phantom{{-}}\underline{0.807} \\
			\phantom{{-}}0.001 \\
			\phantom{{-}}0.073 \\
			\phantom{\vdots} \\
		\end{matrix*}
	}_{\substack{\text{{ABC ETF}} \\ \text{{returns}} }}
	\left.
	\begin{matrix*}[c]
		\phantom{\vdots} \\
		\phantom{.} \\
		\phantom{.} \\
		\phantom{.} \\
		\phantom{.} \\
		\phantom{.} \\
		\phantom{\vdots} 
	\end{matrix*}
	\right]
	=
		\left[
		\begin{matrix*}[c]
			\phantom{\vdots} \\
			\phantom{.} \\ % 1
			\phantom{.} \\ % 2
			\phantom{.} \\ % 3
			\phantom{.} \\ % 4
			\phantom{.} \\ % 5
			\phantom{\vdots} \\
		\end{matrix*}
		\right.
		\underbrace{
		\begin{matrix*}[c]
			\vdots \\
			-0.002 \\
			\phantom{-}0.003 \\
			-0.807 \\
			\phantom{-}0.614 \\
			\phantom{-}0.189 \\
			\vdots \\
		\end{matrix*}
	}_{\substack{\text{{Return}} \\ \text{{spreads}} }}
	\left.
	\begin{matrix*}[c]
		\phantom{\vdots} \\
		\phantom{.} \\ % 1
		\phantom{.} \\ % 2
		\phantom{.} \\ % 3
		\phantom{.} \\ % 4
		\phantom{.} \\ % 5
		\phantom{\vdots} \\
	\end{matrix*}
	\right]
	.
\end{align}
% price. 
In the first two periods, the returns to the synthetic index portfolio are almost the same as the returns to the ETF, producing only a small deviation on the right-hand side. 
In the third period, the ETF price jumps by $0.807$\%, while the prices of the individual stocks do not move much, leading to a large negative spread. 
In the fourth and fifth periods, the prices of the portfolio of individual stocks catch up to the ETF jump, leading to large positive spreads. 
% in the fourth period and a smaller positive spread in the fifth period. 
% which becomes smaller. 

This stylized example captures the central problem in the analysis of common jumps on a fine sampling grid. 
If news reached the entire market instantly, was interpreted homogeneously, and trading were continuous, jumps in a group of stocks should presumably occur simultaneously with the ETF index jump and the spreads should be small and random. Sluggish price changes lead to 
% delays in observed 
asynchronous 
jumps, however, and the spread temporarily expands and contracts again.

\subsubsection{Decomposition of the spread}
\label{sssecEventWindow}

To isolate the effect of 
% asynchronicity in the impoundment of news 
asynchronous jumps 
on the spread, we break up the return of synthetic stock index portfolio $\Delta^n_i S$ into its discontinuous and continuous part: 
\begin{align}
	\label{eqSdisentangle}
	\Delta^n_i S 
	:= 
	\sum_{k = 1}^p {w}_{k,i\Delta_n} \Delta^n_i {Y}_k
	= 
	\sum_{k = 1}^p  {w}_{k,i\Delta_n} \Delta^n_i {J}_k
	+
	\sum_{k = 1}^p  {w}_{k,i\Delta_n}  \Delta^n_i {C}_k. 
\end{align}
Jump tests \citep[like][]{lee2007jumps} flag some large observed returns 	as being jumps. 
We then classify the observed returns as either discontinuous or continuous: 
\begin{align}
	\label{eqClassifyJumps}
	\Delta^n_i {Y}_k 
	&= 
	\Delta^n_i {J}_k + \Delta^n_i {C}_k 
	\, \, \text{and} \, \, 
	\Delta^n_i {Z}^{\phantom{k}} 
	= 
	\Delta^n_i \widetilde{J}^{\phantom{k}}
	+ 
	\Delta^n_i \widetilde{C}^{\phantom{k}},
\end{align}
in which $\Delta^n_i {Y}_k$, for $k = 1, ..., p$, the $i$th return of the one-dimensional observed stock log price process, 
$Y_k$, 
is the sum of 
a discontinuous stock return $\Delta^n_i {J}_k := \Delta^n_i {Y}_k \, \cdot I(\text{Jump}_{i\Delta_n})$
and 
a continuous stock return $\Delta^n_i {C}_k := \Delta^n_i {Y}_k \, \cdot I(\text{No Jump}_{i\Delta_n})$, in which $I(\cdot)$ is an indicator function. 
The continuous and sparse jump return vectors are mutually exclusive.
A similar classification applies to the ETF return $\Delta^n_i {Z}$. 

The return spread \eqref{eqPriceSpread} now equals a linear combination of the weighted stock jumps, the weighted continuous stock returns and the ETF returns: 
\begin{align}
	\label{eqMispricingDecomposition}
	\delta^r_{i\Delta_n}
	\stackrel{(\ref{eqPriceSpread})}{:=}
	\Delta^n_i S
	- 
	\Delta^n_i Z 
	\stackrel{(\ref{eqSdisentangle})}{=}
	\sum_{k = 1}^p 
	\underbrace{{w}_{k,i\Delta_n} \Delta^n_i {J}_k}_{\substack{\text{Weighted} \\ \text{discontinuous} \\ \text{stock returns}}}
	+ \, 
	\underbrace{\sum_{k = 1}^p  \underbrace{{w}_{k,i\Delta_n}  \Delta^n_i {C}_k}_{\substack{\text{Weighted} \\ \text{continuous} \\ \text{stock returns}}} 
		- 
		\underbrace{\Delta^n_i Z}_\text{ETF returns}}_{\text{Target}}.
\end{align}
We want to synchronize the individual discontinuous jumps (the first element) with the target (the second element) to minimize the return spreads on the event window.
If both the stock jumps and the ETF impound news at the same time -- the stock jumps offset the target -- the spread in returns should be small, containing only microstructure noise.

\subsubsection{Constructing the jump-event matrix}
\label{sssecConstructJumpEvent}

To synchronize jumps within an event window, we collect the jump vectors of the individual stocks within  a window of observations: 
\begin{align}
	\label{eqJumpVector}
	[\Delta^n_i {J}_k]_{i \in \mathcal{W}_n}, \, \, \text{with} \, 
	k = 1, ..., p, 
\end{align}
in which $\Delta^n_i {J}_{k}$ is the vector of jump returns for stock $k$, with $k = 1, ..., p$, and $\mathcal{W}_n := [{I_1 \Delta_n},{I_2\Delta_n}]$ is an event window of size $h \equiv {I_2\Delta_n} - {I_1 \Delta_n} + 1$. 
For the empirical application in Section \ref{secEmpirics}, we use an event window from five minutes before to five minutes after the ETF jump, $I_1 \Delta_n = (i^* - 5)\Delta_n$ and $I_2 \Delta_n = (i^* + 5)\Delta_n$. 

To help us rearrange stock jumps, we create an $h \times q$ jump-event matrix $J_n$, that % contain the return spreads 
is an easier-to-handle 
% easier to adjust (?) niet het woord dat 'k zoek. 
representation of the decomposition in 
(\ref{eqMispricingDecomposition}), 
where $h$ is the number of periods in the window around the ETF jump and $q-1$ is the number of jumps in individual stocks, where some stocks might jump more than once or not at all: 
\begin{align}
	\label{eqJumpMatrix}
	{J}_n = (\gamma_{il})
	:= 
	\Big[
	\underbrace{
		{w}_{i\Delta_n} \Delta^n_i {J}
	}_{\substack{\text{Weighted} \\ \text{discontinuous} \\ \text{stock returns}}}, \, 
	\underbrace{T_{i\Delta_n}}_{\text{Target}}
	\Big ]_{i \in \mathcal{W}_n}, 
\end{align}
in which $i = 1, ..., h$ and $l = 1, ..., q$.
The first $q-1$ columns consist of the vectors  of weighted stock discontinuous returns 
${w}_{i\Delta_n} \Delta^n_i {J} := ({w}_{1,i\Delta_n} \Delta^n_i {J}_1, ..., {w}_{q-1,i\Delta_n} \Delta^n_i {J}_{q-1})$ sampled within a window of $h$ observations around the ETF jump. 
The weighted jump vectors ${w}_{i\Delta_n} \Delta^n_i {J}$ 
consist of the $p$ stock jump vectors $\Delta^n_i {J}_k$, for $k = 1, ..., p$ in (\ref{eqJumpVector}), 
but we reorganize the stock jump vectors. If a stock jumps multiple times, 
as does stock A, separate jumps appear in different columns. 
Each jump vector contains one and only one 
non-zero element. 
We exclude the jump vectors for which the stock does not jump, as in the case of stock C. 
The $q$th column is the target vector, $T_{i\Delta_n} := (\sum_{k = 1}^p  {w}_{k,i\Delta_n}  \Delta^n_i {C}_k)  - \Delta^n_i Z$, which is the difference between the continuous returns of the synthetic stock index portfolio, 
$\sum_{k = 1}^p  {w}_{k,i\Delta_n} \Delta^n_i {C}_k$, 
and the ETF returns $\Delta^n_i Z$.  
The elements of the target column cannot be moved, while the stock jumps are the moving parts of the jump-event matrix. 

Spreads are linear combinations of the stock jumps, the stock's continuous returns and the ETF returns (\ref{eqMispricingDecomposition}). They are the row-sums of the jump-event matrix: 
% Row-sums of the jump-event matrix \eqref{eqJumpMatrix} equal the return spreads  (\ref{eqMispricingDecomposition}) across the event window: 
\begin{align}
	\label{eqRowSums}
	J_n^+ 
	:= 
	% J_n \times {1}_q
    % \sum_{m = 1}^{q+1} (\gamma_{im}) 
    % Why q + 1? 
    \sum_{m = 1}^{q} (\gamma_{im}) 
	= [\delta^r_{i\Delta_n}]_{i \in \mathcal{W}_n}.
\end{align}
% in which ${1}_q$ is a column-vector of ones. 

The jump-event matrix \eqref{eqJumpMatrix} in our example looks like: 
\begin{align*}
	{J}_n 
	= 
	\left[
	\begin{matrix*}[c]
		\phantom{.} \\
		\phantom{.} \\
		\phantom{.} \\
		\phantom{.} \\
		\phantom{.} \\
	\end{matrix*}
	\right.
	\underbrace{
		\begin{matrix*}[r]
			0.000 						& 0.000  								& 0.000  \\
			0.000 						& 0.000  								& 0.000   \\
			0.000 						& 0.000  								& 0.000  \\
			\underline{0.210} & 0.000  								& \underline{0.400}  \\
			0.000 						& \underline{0.217}  		& 0.000   \\
		\end{matrix*}
	}_{\substack{\text{Weighted} \\ \text{discontinuous} \\ \text{stock returns}}}
	% white space. the minuses are too close
	\begin{matrix*}[c]
		\phantom{.} \\
		\phantom{.} \\
		\phantom{.} \\
		\phantom{.} \\
		\phantom{.} \\
	\end{matrix*}
	\underbrace{
		\begin{matrix*}[c]
			-0.002 \\
			\phantom{-}0.003 \\
			-0.807 \\
			\phantom{-}0.004 \\
			-0.028 \\
		\end{matrix*}
	}_{\text{Target}}
	\left.
	\begin{matrix*}[c]
		\phantom{.} \\
		\phantom{.} \\
		\phantom{.} \\
		\phantom{.} \\
		\phantom{.} \\
	\end{matrix*}
	\right]
	, 
	\text{with row-sums}  \,  
	J_n^+ = 
	\begin{bmatrix*}[c]
		-0.002  \\ \phantom{-}0.003 \\ -0.807 \\ \phantom{-}0.614 \\ \phantom{-}0.189
	\end{bmatrix*}.
\end{align*}
The first three columns of the jump-event matrix correspond to the individual stock jumps across an event window from two minutes before to two minutes after the ETF jump. 
Note that the number of columns in the first block corresponds to the number of identified jumps in individual stocks, not the number of individual stocks. 
The first two columns contain the gradual jumps of Stock A and the third column contains the delayed jump of stock B. 
The stock jump sizes are generally weighted according to their shares of
the index, but this example uses an equally weighted index for simplicity.
The fourth column of the jump-event matrix is a target vector that contains the  difference between the continuous returns of the stocks and the ETF returns. 
The return spread is the sum of the row-sums of the weighted discontinuous stock returns and the target column, \textit{i.e.} the row-sums of the jump-event matrix (\ref{eqRowSums}). 
As before (in \ref{secMispricing}), there is a negative and a positive spike in the return spread in a small window around the ETF jump in period $3$, because stocks A and B do not jump until periods $4$ and $5$.

\subsection{Rearranging the elements within the jump-event matrix}

After decomposing the stock returns into jump and non-jump returns, 
we can rearrange the stock jumps in the jump-event matrix to offset the target column. 
% , thereby minimizing the variance of the return spreads and recovering the efficient common jump. 
% Rearranging stock jumps within the jump-event matrix moves the jumps in time and changes the return spreads. 

\subsubsection{Permutations and the return spread after rearrangement}
\label{sssecPermutations}

There are different choice options for the rearrangements, making it a combinatorial optimization problem.
% Synchronizing the stock jumps in a jump-event matrix is a combinatorial problem.
We seek to rearrange the stock jumps in the jump-event matrix 
% the $h$ elements in each of weighted stock jump vectors 
(each individual column in the first block of columns in the jump-event matrix) 
to offset the elements in the target vector (the last column in the jump-event matrix), 
which minimizes
the variability of return spreads (the row-sums of the jump-event matrix). 

A rearrangement (of a particular column in the jump-event matrix) is defined by a  permutation ${\pi}_l$ of its $h$ elements: 
$\pi_l: \{1, ..., h\} \rightarrow \{1, ..., h \}$, with $l = 1, ..., q$. The permutation ${\pi}_l$ is represented 
compactly 
by
a vector mapping the original row order into a new row order:
\begin{align}
	\label{eqPermutation}
	\pi_l \equiv 
	\begin{pmatrix}
		% 1 & 2 & ... & h_n \\
		\pi_l (1) & \pi_l (2) & ... & \pi_l (h)
	\end{pmatrix}.
\end{align}
The vector of $q$ permutations ${\pi} := ({\pi}_1, ..., {\pi}_q)$ collects the rearrangements of all columns. 
Note that the $q$th column, the target, is fixed. 
We do not swap any elements in the $q$th column of the jump-event matrix. 

Each rearrangement of an observed jump-event matrix (\ref{eqJumpMatrix}) yields a new (``rearranged") jump-event matrix: 
\begin{align}
	{J}_n 
	:= 
	\Big[
	\underbrace{
		{w}_{i\Delta_n} \Delta^n_i {J}
		% {w}^1_{i\Delta_n} \Delta^n_i {J}^1, ..., {w}^p_{i\Delta_n} \Delta^n_i {J}^p
	}_{\substack{\text{Weighted} \\ \text{discontinuous} \\ \text{stock returns}}}, \, 
	\underbrace{T_{i\Delta_n}}_{\text{Target}}
	\Big ]_{i \in \mathcal{W}_n}
	%
	% \, \,  \text{with} \, \, 
	% {w}_{i\Delta_n} \Delta^n_i {J} 
	\xrightarrow[]{\text{Rearrangement}} \, 
	% {J}_n^\pi := \left[
	%{w}^1_{i\Delta_n} \Delta^n_i {J}^{1,\pi}, {w}^2_{i\Delta_n} %\Delta^n_i {J}^{2,\pi}, ..., {w}^p_{i\Delta_n} \Delta^n_i %{J}^{p,\pi},
	%T_{i\Delta_n}
	%\right]_{i \in \mathcal{W}_n}. \\ 
	%
	% \left[{w}_{i\Delta_n} \Delta^n_i {J}^\pi, \, 
	% T_{i\Delta_n}
	% \right]_{i \in \mathcal{W}_n} := {J}_n^\pi, 
	\Big[
	\underbrace{
		{w}_{i\Delta_n} \Delta^n_i {J}^\pi
		% {w}^1_{i\Delta_n} \Delta^n_i {J}^1, ..., {w}^p_{i\Delta_n} \Delta^n_i {J}^p
	}_{\substack{\text{Rearranged} \\ \text{weighted} \\ \text{discontinuous} \\ \text{stock returns}}}, \, 
	\underbrace{T_{i\Delta_n}}_{\text{Target}}
	\Big ]_{i \in \mathcal{W}_n} 
	:= {J}_n^\pi,
\end{align}
\sloppy
in which
% $J_n = (\gamma_{il})$ is the observed jump-event matrix (\ref{eqJumpMatrix}), 
$J^\pi_n = (\gamma^\pi_{il})$ is the rearranged jump-event matrix, 
${w}_{i\Delta_n} \Delta^n_i {J}^\pi := 
({w}_{1,i\Delta_n} \Delta^n_i {J}^{\pi}_1, ..., {w}_{q-1,i\Delta_n} \Delta^n_i {J}^{\pi}_{q-1})$ is the vector of \textit{rearranged} weighted stock % discontinuous 
jump 
returns.
% 

% \sloppy
The row-sums of the rearranged jump-event matrix $J_n^\pi$, 
\textit{i.e.} the corresponding return spreads, 
are % now a function of the arrangement of the stock jumps:
expressed as a function of the arrangement (and timing) of the stock jumps: 
\begin{align*}
	J_n^{\pi,+} := 
	% J^\pi_n \times 1_q 
	% = 
    % \sum_{m = 1}^{q+1} (\gamma^\pi_{im}).
    % Why q+1?
    \sum_{m = 1}^{q} (\gamma^\pi_{im})
	% [\delta^{r,\pi}_{i\Delta_n}]_{i \in \mathcal{W}_n}.
    \end{align*}
 
For example, consider the following permutation ${\pi}_1$ \eqref{eqPermutation} of the first column of the jump-event matrix, swapping the 3rd and the 4th observation: 
\begin{align*}
	\pi_1 = 
	\begin{pmatrix}
		% 1 & 2 & 3 & 4 & 5 \\
		1 & 2 & \underline{4} & \underline{3} & 5
	\end{pmatrix}. 
\end{align*} 
This swap rearranges the jump-event matrix, switching the 3rd and 4th rows of the first column: 
\begin{align*}
	{J}_n^\pi = 
	\left[
	\begin{matrix}
		\phantom{.} \\
		\phantom{.} \\
		\phantom{.} \\
		\phantom{.} \\
		\phantom{.} \\
	\end{matrix}
	\right.
	\underbrace{
		\begin{matrix}
			0.000 								& 0.000 						& 0.000  \\
			0.000 								& 0.000 						& 0.000  \\
			\underline{0.210}			& 0.000 						& 0.000  \\
			0.000									& 0.000 						& \underline{0.400}  \\
			0.000 								& \underline{0.217} & 0.000  \\
		\end{matrix}
	}_{\substack{\text{Rearranged} \\ \text{weighted} \\ \text{discontinuous} \\ \text{stock returns}}}
	% white space. the minuses are too close
	\begin{matrix}
		\phantom{.} \\
		\phantom{.} \\
		\phantom{.} \\
		\phantom{.} \\
		\phantom{.} \\
	\end{matrix}
	\underbrace{
		\begin{matrix}
			-0.002 \\
			\phantom{-}0.003 \\
			-0.807 \\
			\phantom{-}0.004 \\
			-0.028 \\
		\end{matrix}
	}_{\text{Target}}
	\left.
	\begin{matrix}
		\phantom{.} \\
		\phantom{.} \\
		\phantom{.} \\
		\phantom{.} \\
		\phantom{.} \\
	\end{matrix}
	\right]
	, 
	\text{with row-sums}  \, \,   
	J_n^{\pi,+} = % \sum_{m = 1}^{q+1} (\gamma_{im}) = 
	\begin{bmatrix}
		-0.002 \\ 
		\phantom{-}0.003  \\ 
		-0.597 \\ 
		\phantom{-}0.404  \\
		\phantom{-}0.189 
	\end{bmatrix}, 
\end{align*}
which shifts a weighted stock jump of stock A one period back in time (and a zero forward in time).  
The permutation $\pi_1$ also changes the third and fourth row-sum of the jump-event matrix. 
The variability in the % return spreads 
row-sums 
is slightly smaller after this rearrangement, because the ETF jump also occurs in the third observation.

\subsection{The best rearrangement of the jump-event matrix}

Under the assumption that the latent prices of the price series of the stocks and the ETF move in lockstep, 
the best rearrangement moves jumps in time to minimize the variability of the return spreads. 
% the deviations of the synthetic portfolio returns from the ETF returns. 
This reduction in variability is known as ``flattening".\footnote{%
    Flattening the row-sums of the jump-event matrix means that the stochastic variables in the separate columns of the jump-event matrix should be completely mixable. 
    Mathematically, a $q$-dimensional distribution function $F(Q_1, ..., Q_q)$ on $\mathbb{R}$ is {$q$-completely mixable} if there exist $q$ random variables $Q_1, ..., Q_q$ identically distributed as $F$ such that: 
    \begin{align*}
    	\text{Prob} \left(Q_{1} + Q_2 + ... + Q_q = \text{constant}\right) = 1.
    \end{align*}
    That is, the sums of $q$ random variables drawn from a $q$-completely mixable distribution should approximate a constant.
    A completely mixable dependence structure minimizes the variance of the sum of the random variables with given marginal distributions. 
    In a discrete case, like the jump-event matrix which consists of realizations of random variables, we look for a particular ordering in each of the columns, such that the row-sums approximate a constant.}

The optimization problem that minimizes the variance of the return spreads is a combinatorial problem that can be expressed as follows:
\begin{align}
	\label{eqV}
	\min_\pi V \left( J_n^{\pi,+} \right), \, \, \text{with} \, \, J_n^{\pi,+} := \sum_{m = 1}^{q} (\gamma^\pi_{im}),
\end{align}
in which the row-sums $J_n^{\pi,+}$ are return spreads expressed as a function of the arrangement of stock jumps,
$V(\cdot)$ is a scalar-valued function that measures the variability, 
\textit{e.g.}, the range, 
of the vector of row-sums. 
% Minimizing, minimizes the deviations of the synthetic portfolio returns from the ETF returns. 

The combinatorial optimization problem (\ref{eqV}) is rooted in the pioneering work of \citet{puccetti2012computation} and \citet{embrechts2013model} on rearrangements and the Rearrangement Algorithm; 
looping over each column of a matrix to order it oppositely to the sum of the other columns (see Appendix \ref{ssecRA} for an example). 
This algorithm is best known as an actuarial tool to bound portfolio risk, but it also has applications in other disciplines, such as operations research \citep[see \textit{e.g.},][]{boudt2018block}. 
The algorithm can propose a best rearrangement of the jump-event matrix \eqref{eqV} but it does not constrain the type of rearrangements that take place. For example, it can move jumps either forward or backward in time to any point in the window. 

To constrain the procedure from economically implausible rearrangements, we introduce the Rearrangement Linear Program (RLP) that is well suited to choose arguments which minimize an objective function 
% \ref{eqV}, 
(\ref{ssecDecisionVariable}), subject to linear constraints \eqref{sssecConstraint} and penalties \eqref{sssecPenalties}. 
\if1\comment
\blue{The next parts (Objective function and Constraints) are badly explained and are in desperate need of a rewrite. Let's first find a decent empirical application and then go for an overhaul of the next subsections.}
% Hoe hebben we dit gedaan in de slides? 
% Simplex stuff. 
% https://en.wikipedia.org/wiki/Linear_programming
\fi

\subsubsection{Arguments and objective function}
\label{ssecDecisionVariable}

There are two types of arguments to the solution function: 1) permutation matrices and 2) an unknown interval within which all the row-sums lie. 
% unknown range; the interval of the largest and smallest row-sum. 

\subsubsection*{Permutation matrices}

% The RLP chooses permutation matrices to rearrange jumps. 
% and minimize the  objective function. 
To rearrange (i.e., permute) a column of the jump-event matrix, we premultiply a column by a permutation matrix. 
% To constrain the rearrangements, 
We rely on the column representation of a permutation matrix. 
Each permutation matrix permutes one column of the jump event matrix, so a solution will include $q$ permutation matrices -- 4 columns in the jump-event matrix means there are 4 permutation matrices. 

An $h \times h$ permutation matrix permutes the columns of the identity matrix $I_{h}$ to express a permutation $\pi_l$ \eqref{eqPermutation}: 
\begin{align}
	\label{eqPermuMatrix}
	P_{\pi_l} 
	% = (p_{i,k}) 
	= (p_{ii'}) 
	= 
	\begin{bmatrix} 
		p_{11} &   p_{12} &  \hdots & p_{1h}  \\ 
		p_{21} &   p_{22} &   \hdots & p_{2h} \\ 
		\vdots &          &   \ddots &  \\ 
		p_{h1} &   p_{h2} &     \hdots & p_{hh}
	\end{bmatrix}
	= 
	\begin{bmatrix}
		\bm{e}_{\pi_l(1)} \\
		\bm{e}_{\pi_l(2)} \\
		\vdots \\
		\bm{e}_{\pi_l(n)} \\
	\end{bmatrix}.
\end{align}
For each $i$, 
$p_{ii'}$ is $1$ if $i' = \pi_l (i)$ and is $0$ otherwise. 
The entries of the $i$th row are all zero except for a $1$ that appears in column $\pi_l (i)$. 
A standard basis vector, $\bm{e}_{i'}$, denotes a row-vector of length $h$ with a $1$ on position $i'$ and a $0$ on every other position. 

We rely on this representation because a permutation matrix \eqref{eqPermuMatrix} can track how far the ones 
deviate from the diagonal, which permits penalties for movement. 
For example, we can impose a maximum distance from the diagonal to not let the jumps stray too far in the event window (see Section \ref{sssecPenalties} for further elaboration). 

% The permutation (matrix) below would switch the 3rd and 4th rows of the 1st column of a jump-event matrix
The permutation in the example of Section \ref{sssecPermutations}, $\pi_1 = 
\begin{pmatrix}
	1 & 2 & \underline{4} & \underline{3} & 5 
\end{pmatrix}$, 
switches  the 3rd and 4th rows of the 1st column of a jump-event matrix, and 
is equivalent to the following permutation matrix:  
\begin{align*}
	P_{\pi_1} = 
	\begin{bmatrix}
		\bm{e}_{\pi_1(1)} \\
		\bm{e}_{\pi_1(2)} \\
		\vdots \\
		\bm{e}_{\pi_1(n)} \\
	\end{bmatrix}
	= 
	\begin{bmatrix}
		\bm{e}_{1} \\
		\bm{e}_{2} \\
		\bm{e}_{4} \\
		\bm{e}_{3} \\
		\bm{e}_{5}
	\end{bmatrix}
	= 
	\begin{bmatrix}
		1 & 0 & 0 & 0 & 0 \\
		0 & 1 & 0 & 0 & 0 \\
		0 & 0 & 0 & \underline{1} & 0 \\
		0 & 0 & \underline{1} & 0 & 0 \\
		0 & 0 & 0 & 0 & 1 
	\end{bmatrix},
\end{align*}
in which column $i'$ of the $I_5$ identity matrix now appears as the column $\pi(i')$ of $P_{\pi_1}$.
The changes occur in the vertical, $i$, dimension. 
Upward moves in the permutation matrix are backward moves in time. Downward moves in the permutation matrix are forward moves in time.
The fourth element, which is on position $(4,4)$ in $I_5$, shifts one spot backward in time 
by shifting one step upward in the permutation matrix. 
The third element, which is on position $(3,3)$ in $I_5$, shifts one step forward in time by shifting one step downward in the permutation matrix. 

We have a permutation matrix for each of the $q$ columns in the jump-event matrix. 
We concatenate the permutation matrices in a co-permutation matrix of dimension $h \times (h q)$: 
\begin{align}
	\label{eqCoPermuMatrix}
	\Pi 
	% P_\pi 
	= \left(p_{lii'}\right) 
	= \left[P_{\pi_1}, P_{\pi_2}, ..., 				P_{\pi_q}\right],
\end{align}
with $l = 1, ..., q$, $i,i' = 1, ..., h$ 
and 
$P_{\pi_{1}}$
short notation for the $h \times h$ permutation matrix for the first column of the jump-event matrix. 

Premultipying the (vectorized) jump-event matrix by the co-permutation matrix produces the vector of return spreads:
% The permutation matrices change the row-sums in the objective function \eqref{eqV}. 
% To minimize the variability of the row-sums \eqref{eqV}, we minimize the {range} of the row-sums. 
% For each rearrangement of the jump-event matrix,  $J_n = (\gamma_{il})$, with $i = 1, ..., h$ and $l = 1, ..., q$, we can 
% rewrite the row-sums as a linear function of the co-permutation matrix \eqref{eqCoPermuMatrix}: 
\begin{align}
	% \label{Wmat}
	\label{eqLPRow-sums}
	J_n^{\pi,+} = 
	% P^\pi \, 
	\Pi
	\times 
	% \text{vec}(X) % X is dus diegene die niet herschikt is. 
	\text{vec}(J_n),
	% = S_\pi
\end{align}
or, equivalently: 
\begin{align*}
	\begin{bmatrix}
		J^{\pi,+}_{n,1} \\
		J^{\pi,+}_{n,2} \\
		J^{\pi,+}_{n,3} \\
		\vdots \\
		J^{\pi,+}_{n,h}
	\end{bmatrix} 
	\hspace{-.1cm}
	= 
	\hspace{-.1cm}
	\left[
	\begin{array}{cccc | cccc | c | cccc}
		p_{111} & p_{112} & \hdots & p_{11h} & p_{211} & p_{212} & \hdots & p_{21h} & 	   \hdots & p_{q11} & p_{q12} & \hdots & p_{q1h} \\
		p_{121} & p_{122} & \hdots & p_{12h} & p_{221} & p_{222} & \hdots & p_{22h} & 	   \hdots & p_{q21} & p_{q22} & \hdots & p_{q2h} \\
		\vdots &   & \ddots & \vdots & 	 	\vdots &   & \ddots & \vdots &   	 	     & \vdots &   & \ddots & \vdots	 \\
		p_{1h1} & p_{1h2} & \hdots & p_{1hh} & p_{2h1} & p_{2h2} & \hdots & p_{2hh} & \hdots & p_{qh1} & p_{qh2} & \hdots & p_{qhh} 
	\end{array}
	\right]
	\hspace{-.1cm}
	\times  
	\hspace{-.1cm}
	\begin{bmatrix}
		\gamma_{11} \\
		\gamma_{21} \\
		\vdots \\
		\underline{\gamma_{h1}} \\
		\vdots \\
		\gamma_{hq}
	\end{bmatrix},
\end{align*}
in which $\text{vec}(J_n)$, the vectorized version of the observed jump-event matrix $J_n = (\gamma_{il})$, with $i = 1, ..., h$ and $l = 1, ..., q$, 
is a stacked column vector of dimension $hq \times 1$. 
The result of the matrix product in \eqref{eqLPRow-sums} is a $h \times 1$ column  vector, including the row-sums of the rearranged jump-event matrix (the left side of the equal sign). 

% Add example again? 

\subsubsection*{Unknown interval}

The RLP chooses a co-permutation matrix that minimizes the range of the row-sum. The range is the difference between the maximum and the minimum order statistic of the row-sums: %  for a particular rearrangement: 
\begin{align}
	\label{eqRange}
	R = J^{\pi,+}_{n,(h)} - J^{\pi,+}_{n,(1)}.
\end{align}
in which the subscript $(i)$ enclosed in parentheses indicates the $i$th order statistic of the sample: the smallest and a largest row-sum for a particular arrangement. 

The positions of the smallest and largest row-sums are unknown upfront. In order to express this objective function within the RLP we 
% cannot extract it directly from the % co-permutation matrix \eqref{eqCoPermuMatrix}, 
% matrix product in \eqref{eqLPRow-sums}. 
slightly deviate from the standard canonical forms of linear programs -- 
% because we cannot apply mathematical operations like the minimum and maximum on the row-sums within the linear program. 
the objective function within a linear program is typically an affine function of its arguments.\footnote{Linear programs are problems that can be expressed in canonical form as: ``Find a vector $\bm{x}$ that minimizes $\bm{c}^\top \bm{x}$, with $\bm{c}$ a given vector, subject to some constraints on $\bm{x}$."}
% er bestaat geen c... x is onbekend. dus hoe zou onze c eruit zien.. Zie slides. 
% One would think that  
% Intuitively, 
% Unfortunately, the range \eqref{eqRange}, which includes a maximum and a minimum, is not a direct function of the matrix product in \eqref{eqLPRow-sums}. 
% one would think that 
% the only % lever
% argument in the linear program is 
The co-permutation matrix extracts the row-sum in each row in the matrix product \eqref{eqLPRow-sums}, 
\textit{i.e.}, $J^{\pi,+}_{n,1}, J^{\pi,+}_{n,2}, ..., J^{\pi,+}_{n,h}$ (without brackets), 
but there is no possible choice of the co-permutation matrix which could ever produce the maximum or minimum of the row-sums in \eqref{eqRange}, i.e., $J^{\pi,+}_{n,(1)}$ and $J^{\pi,+}_{n,(h)}$ (with brackets). 
% at least not directly.
% \footnote{For example, by setting some elements 
% some blocks
% in the co-permutation matrix to zero and changing the signs of the first block, the matrix product in \eqref{eqLPRow-sums} could minimize the difference between specific row-sums, like the the difference between the row-sum in the last and the first row, $J^{\pi,+}_{n,h} - J^{\pi,+}_{n,1}$, but never the range $J^{\pi,+}_{n,(h)} - J^{\pi,+}_{n,(1)}$. This would only be be helpful in flattening a comonotone matrix in which each column is a sequence from 1 to 5  
%	 (see \textit{e.g.} \citeauthor{embrechts2013model}, \citeyear{embrechts2013model}), 
%	but not a jump-event matrix for which the row-sum in the first row is not necessarily the smallest number.
% } 
% Therefore, we treat it as an additional argument in the optimization problem. 

We get the appropriate objective function indirectly, by minimizing an unknown interval within which all the row-sums lie: 
\begin{align}
	\label{minRange}
	\text{Find } \Pi,L,U \text{ that minimizes } 
	U - L, 
\end{align}
in which $\Pi$ is the co-permutation matrix, which defines arrangement of the elements in the  rearranged jump-event matrix, 
$U$ is the upper boundary of an unknown interval and 
$L$ is the lower boundary of an unknown interval. 

The program chooses candidates for both types decision variables -- the permutation matrices and the lower and upper boundary of the unknown interval -- 
to minimize the range of that interval \eqref{minRange}.
The MILP 	chooses the elements (0,1) 	for  the permutation matrices and 	continuous values for the 	lower and upper boundary. 
The optimization problem \eqref{minRange}  is therefore a mixed-integer linear program (MILP). 

The optimization in \eqref{minRange} does not minimize the range of the row-sums yet. We must  
define a permutation within a linear program and  
connect the lower and upper boundary, $L$ and $U$ with the co-permutation matrix $\Pi$
by constraining the choices 
of the decision variables.

        \subsubsection{Constraints}
        \label{sssecConstraint}

We constrain the sensible choices of the decision variables using three types of constraints: the ordering constraint, the permutation constraint and a target constraint. 
The combination of these minimal conditions does lead to appropriate rearrangements which minimize the range. 

            \subsubsection*{The ordering constraint}

The unconstrained MILP, as it is defined in \eqref{minRange}, has a lower and upper boundary of an unknown interval both in the argument and in the objective function. 
It can choose any continuous value for the lower boundary $L$ 	(say, a big negative number) and any continuous value for the upper boundary (say, zero) to minimize the difference between these two random numbers. 
The minimization will push the difference towards a big negative number, because these optimal boundaries 	are still unconnected to the co-permutation matrix.

We impose inequality constraints on the boundaries: 
\begin{align}
	% \tag{C.1}
	\label{eqConstrRowSums}
	% J^{\pi,+}_{n,(1)} \leq J^{\pi,+}_{n,i} \leq J^{\pi,+}_{n,(h)},  \, \,  \text{for} \,\,  i = 1, ..., h, 
	% J^{\pi,+}_{n,(1)} 
	L
	\leq J^{\pi,+}_{n,i} \leq 
	% J^{\pi,+}_{n,(h)}, 
	U, 
	 \, \,  \text{for} \,\,  i = 1, ..., h, 
\end{align}
in which the row-sums in the middle are the result of the matrix product in \eqref{eqLPRow-sums} for a particular arrangement of jumps. 
The constraint indirectly defines the smallest possible (the minimum) and largest possible (maximum) row-sum. 
The left inequality in \eqref{eqConstrRowSums}  defines a lower boundary 
  in a set of row-sums ({by definition} each individual row-sum should greater than or equal to the minimum) 
 and the right inequality defines 
 upper boundary 
  in a set of row-sums (by definition each individual row-sum should be less than or equal to the maximum). 
The outer boundaries, 
the lower boundary 
should be less than or equal to the 
upper boundary, 
ensures that the minimum is always smaller than the maximum, as by definition. 

Choosing 
candidate values for the 
	lower and upper boundary \eqref{minRange}
	still result in values which are unconnected to the row-sums. 
	A big negative number for the lower boundary $L$ will  satisfy the constraint \eqref{eqConstrRowSums}: $L$ will  be smaller than any row-sum.
	(And a big positive number for the upper boundary will also satisfy the constraint: $U$ will still be larger than any row-sum.)
But by minimizing the difference of these decision variables, $U - L$, as defined in  \eqref{minRange}, 
	the RLP squeezes the outer values in the constraint \eqref{eqConstrRowSums} % $L$ and $U$ 
    together, 
	as close as possible. (Note that if we would maximize the range, $U - L$, the lower and outer boundary $L$ and $U$ would move away from each other and from the row-sums.)
    The optimal solutions for $L$ and $U$ will then be equal to % become % indirectly defining 
    the smallest and largest row-sum: 
	\begin{align}
	% L \rightarrow J^{\pi,+}_{n,(1)} \, \, \,  \text{and} \, \, \, 
	% U \rightarrow  J^{\pi,+}_{n,(h)},
    L^* = J^{\pi,+}_{n,(1)} \, \, \,  \text{and} \, \, \, 
	U^* =  J^{\pi,+}_{n,(h)},
	\end{align}
    A simple proof by contradiction suffices to confirm this statement. 
    Suppose that $L^*$ is not equal to the smallest row-sum $J^{\pi,+}_{n,(1)}$ or $U^*$ is not equal to the largest row-sum $J^{\pi,+}_{n,(h)}$. In any of those cases, the RLP can still further minimize the objective function $U - L$.
    
\subsubsection*{The permutation constraints}

To define a proper co-permutation matrix \eqref{eqCoPermuMatrix}, we impose equality constraints on the linear program: 
 \begin{align}
 	%% SELECTION
 	\label{eqPermu1}
 	% \tag{C.2a}
 	% \sum_{i=1}^n \sum_{j=1}^d \sum_{k=1}^{n} p_{ijk} = dn
 	% \sum_{i=1}^h \sum_{j=1}^q \sum_{k=1}^{h} p_{ijk} = hq
 	\sum_{l=1}^q \sum_{i=1}^h \sum_{i'=1}^{h} p_{lii'} &= hq, 
 	\\
 	%
 	%% ROW
 	\label{eqPermu2}
 	% \tag{C.2b}
 	% \sum_{k=1}^{n} p_{ijk} = 1,  \, \text{for} \, \, i = 1, ...,n; j = 1, ..., d
 	\sum_{i'=1}^{h} p_{lii'} &= 1,  \, \, \text{for} \, \, i = 1, ...,h\,\,  \text{and} \,\, l = 1, ..., q,
 	\\
 	\label{eqPermu3}
 	%% COLUMN
 	\sum_{i=1}^{h} p_{lii'} &= 1,  \, \, \text{for} \, \, l = 1, ..., q \, \, \text{and} \, \,  i' = 1, ...,h.
 \end{align}
The first equation \eqref{eqPermu1} requires that each permutation matrix (\ref{eqPermuMatrix}) has $h$ ones to select all (exactly $h$) elements in each column of the jump-event matrix, so the sum over all permutation matrices in the co-permutation matrix should equal $hq$. 
The second equation \eqref{eqPermu2} constrains the rows on each permutation matrix to sum to one. 
Suppose that the rows of each permutation matrix do not sum to one, then we could have either multiple or zero elements in the rearranged matrix on a particular position. 
The last equation \eqref{eqPermu3} is a column constraint on the permutation matrix, which guarantees that the same number does not appear twice in a column of the rearranged matrix, even if the rows sum to one. 

        \subsubsection*{The target}

We also impose an equality constraint on the permutation matrix of the last column, so that the RLP does not rearrange the last column of the jump-event matrix, \textit{i.e.} the target: 
\begin{align}
	\label{eqConstrTarget}
	(\text{diag} ( P_{\pi_{q}} ))_i   =  1,  \, \, \text{for} \, \, i = 1, ...,h, 
\end{align}
The permutation \eqref{eqPermuMatrix} matrix corresponding the $q$th column of the jump-event matrix $P_{\pi_{q}}$  should remain an identity matrix $I_n$.

\subsubsection{Penalties}
\label{sssecPenalties}

A linear program is flexible. It allows for the introduction of penalties to prohibit economically implausible rearrangements. For example, we could penalize large 
moves in time with the aid of a distance matrix.
The distance matrix $D_n$ tracks the distance from the diagonal in any permutation matrix $P_{\pi_l}$ \eqref{eqPermuMatrix} of the same size: 
\begin{align}
	\label{eqDistanceMatrix}
	D_n 
	= \left(d_{ii'} \right)
	\left[
	\begin{array}{ccccccc }
		0 & 1 & \cdots & h-2 & h-1 \\
		1 & 0 & \cdots & h-3 & h-2\\
		\vdots & \vdots & \ddots   &  \vdots & \vdots\\
		h-2 & h-3 & \cdots & 0 & 1 \\
		h-1 & h-2 & \cdots & 1 & 0 \\ 
	\end{array}
	\right].
\end{align}
Keeping the elements on the diagonal of a permutation matrix $P_{\pi_l}$  \eqref{eqPermuMatrix}, \textit{i.e.} no moves, results in a zero distance: the diagonal elements in the distance matrix % \eqref{eqDistanceMatrix} 
$D_n$
are zero. 

The distance matrix also allows us to enforce an economic assumption: 
the RLP only allows for a rearrangement of jumps backward in time because we assume that stock prices are sluggish and lag the highly liquid and carefully watched ETF, they do not lead it.
That is, we only permit stock jumps to be moved to an earlier time (upward moves in the permutation matrix $P_{\pi_l}$), not a later time (downward moves in the permutation matrix $P_{\pi_l}$). 
By taking the upper triangular portion of the distance matrix $D_n$,  we can focus on the   backward shifts  in a permutation matrix $P_{\pi_l}$.  
The upper triangular portion of the distance matrix still tracks the backward distance traveled of \textit{all} elements in a column of the jump-event matrix, because the permutation matrix \eqref{eqPermuMatrix} also tracks the rearrangements of the zeros. 
To only track the move of a stock jump, we disable the irrelevant columns of the distance matrix: if the price jumps in period $i = i^*$, we set the columns $i' \neq i^*$ of $D_n$ to zero.

To limit the length of jump moves in our rearrangements, we impose an inequality constraint on a distance metric:  
\begin{align}
	% \tag{P1}
	\label{eqPenalty}
	d(P_{\pi_l}) \leq c,  \, \text{for} \, \, l = 1, ..., q - 1, 
\end{align}
in which $c \geq 0$ is the maximum permitted length of the backward move and $d(P_{\pi_l})$ is the total distance traveled of a jump within the $l$th column of the jump-event matrix.%
    \footnote{The total number of relevant moves for the $l$th column in the jump-event matrix is equal to the following matrix product of a vectorized permutation matrix and a vectorized distance matrix: 
    \begin{align}
    	d(P_{\pi_l}) =  
    	\text{vecr} 
    	\left( P_{\pi_l} \right)
    	\times 
    	\text{vecr} \left( D_n \right)^\top, 
    	\,\,  \text{for} \, \, l = 1, ..., q.
    \end{align}
    The vectorization, $\text{vecr} (\cdot)$ 
    concatenates 
    the rows of a matrix, 
    as opposed to a standard vectorization which stacks the the columns, producing a $1 \times hq$ row-vector. We transpose the second term after vectorization to get a $hq\times 1$ column-vector. 
    The result of this matrix product is a row-wise multiplication of the elements of the two matrices, which equals the total number of relevant shifts.}
We do not constrain the $q$th column, because the RLP keeps the target column fixed in constraint \eqref{eqConstrTarget}. 
It is possible to constrain each stock differently depending on the liquidity of the stock. 

\begin{figure}[htb!]
	\caption{Trace plot: Range of the % return 
		spreads and 
		jump arrival times as a function of the permitted maximum {length of the move}}
	\centering
	
	\includegraphics[width=.9\textwidth,angle = -0]{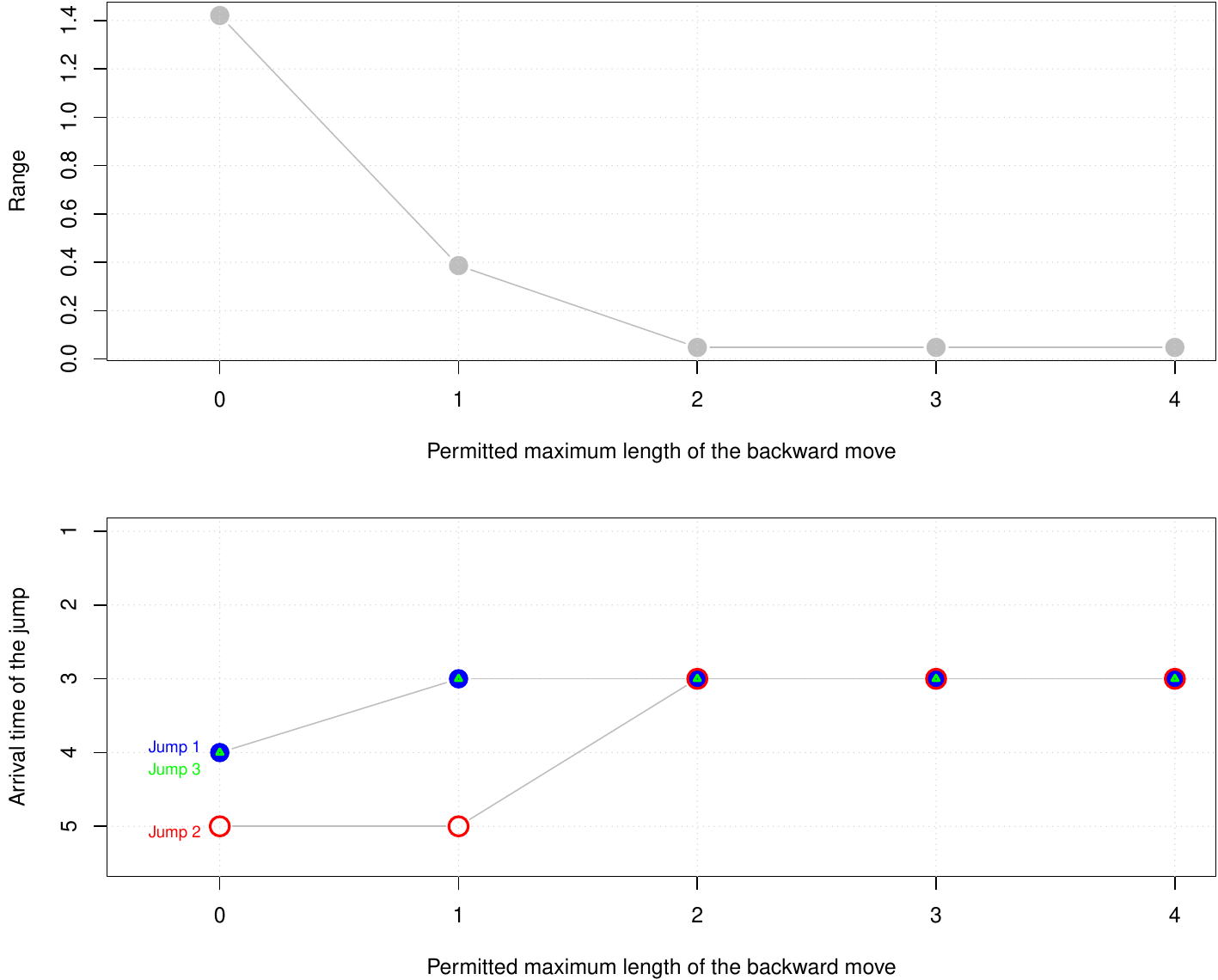}
	\label{fig:RALP-Jump-Asymmetry}
		\begin{minipage}{1.0\linewidth}
			\begin{tablenotes}
				\small
				\item {
					\medskip
					Note: 
					This figure plots the range of the 					return spreads (the row-sums of the jump-event matrix) and the implied jump arrival times 
					as a function of the {length} of permitted backward moves of the jumps in each of the columns.  
					The RLP rearranges the gradual jump of Stock A (Jump 1 in the 4th period and Jump 2 in the 5th period) and the delayed jump of Stock B (Jump 3 in the 4th period). 
					We allow each jump to move backward in time a maximum of zero, one, two, three, four minutes and solve the RLP for each of these five possible constraints.
				}
			\end{tablenotes}
		\end{minipage}
\end{figure}

Figure \ref{fig:RALP-Jump-Asymmetry} shows the range of the spreads, \textit{i.e.}, flatness, and the jump arrival times as a function of the permitted maximum length of the move for each jump in the  stylistic jump-event matrix. 
We allow each jump to move backward in time a maximum of zero, one, two, three, four minutes and solve the RLP for each of these five possible jump-length constraints.
The starting range is what we observe: the gradual jump of Stock A (Jump 1 in the 4th period and Jump 2 in the 5th period) and the delayed jump of Stock B (Jump 3 in the 4th period) with a relatively high range of 
$1.421$. 
The negative slope of the line in the top panel of Figure \ref{fig:RALP-Jump-Asymmetry} shows that allowing more backward shifts flattens the return spreads. 

The bottom panel shows the arrival periods for the jumps as a function of the permitted 
length
of backward moves. With no backward moves, jumps 1 and 3 arrive in period 4 and jump 3 arrives in period 5.  
If we permit one backward move for each jump, the RLP moves jumps 1 and 3 to period 3 while if we permit 2 backward moves, the RLP moves all jumps to period 3. 
The smallest range ($0.048$) occurs with two backward shifts, aligning all the jumps in the third period in the bottom panel. 

Optimally rearranging the jumps, produces the following transformation of the jump-event matrix: 
	\begin{align*}
		{J}_n 
		= 
		\left[
		\begin{matrix*}[c]
			\phantom{.} \\
			\phantom{.} \\
			\phantom{.} \\
			\phantom{.} \\
			\phantom{.} \\
		\end{matrix*}
		\right.
		\underbrace{
			\begin{matrix*}[r]
				0.000 								& 0.000 						& 0.000  \\
				0.000 								& 0.000 						& 0.000  \\
				0.000 								& 0.000 						& 0.000  \\
				\underline{0.210} 		& 0.000 						& \underline{0.400}  \\
				0.000 								& \underline{0.217} & 0.000  \\
			\end{matrix*}
		}_{\substack{\text{Weighted} \\ \text{discontinuous} \\ \text{stock returns}}}
		% white space. the minuses are too close
		\begin{matrix*}[c]
			\phantom{.} \\
			\phantom{.} \\
			\phantom{.} \\
			\phantom{.} \\
			\phantom{.} \\
		\end{matrix*}
		\underbrace{
			\begin{matrix*}[c]
				-0.002 \\
				\phantom{-}0.003 \\
				-0.807 \\
				\phantom{-}0.004 \\
				-0.028 \\
			\end{matrix*}
		}_{\text{Target}}
		\left.
		\begin{matrix*}[c]
			\phantom{.} \\
			\phantom{.} \\
			\phantom{.} \\
			\phantom{.} \\
			\phantom{.} \\
		\end{matrix*}
		\right]
		%%%
		&\xrightarrow[\text{rearrangement}]{\text{The best}}
		%%%
		\left[
		\begin{matrix*}[c]
			\phantom{.} \\
			\phantom{.} \\
			\phantom{.} \\
			\phantom{.} \\
			\phantom{.} \\
		\end{matrix*}
		\right.
		\underbrace{
			\begin{matrix*}[r]
				0.000 								& 0.000 						& 0.000  \\
				0.000 								& 0.000 						& 0.000  \\
				\underline{0.210} 		& \underline{0.217} & \underline{0.400}  \\
				0.000 								& 0.000 						& 0.000  \\
				0.000 								& 0.000 						& 0.000  \\
			\end{matrix*}
		}_{\substack{\text{Rearranged} \\ \text{weighted} \\ \text{discontinuous} \\ \text{stock returns}}}
		% white space. the minuses are too close
		\begin{matrix*}[c]
			\phantom{.} \\
			\phantom{.} \\
			\phantom{.} \\
			\phantom{.} \\
			\phantom{.} \\
		\end{matrix*}
		\underbrace{
			\begin{matrix*}[c]
				-0.002 \\
				\phantom{-}0.003 \\
				-0.807 \\
				\phantom{-}0.004 \\
				-0.028 \\
			\end{matrix*}
		}_{\text{Target}}
		\left.
		\begin{matrix*}[c]
			\phantom{.} \\
			\phantom{.} \\
			\phantom{.} \\
			\phantom{.} \\
			\phantom{.} \\
		\end{matrix*}
		\right]
		= {J}_n^\pi. 
	\end{align*}

\begin{figure}[p]
	\caption{The best rearrangement approximately recovers the efficient stock jumps}
	\centering
	
	\includegraphics[width=.9\textwidth,angle = -0]{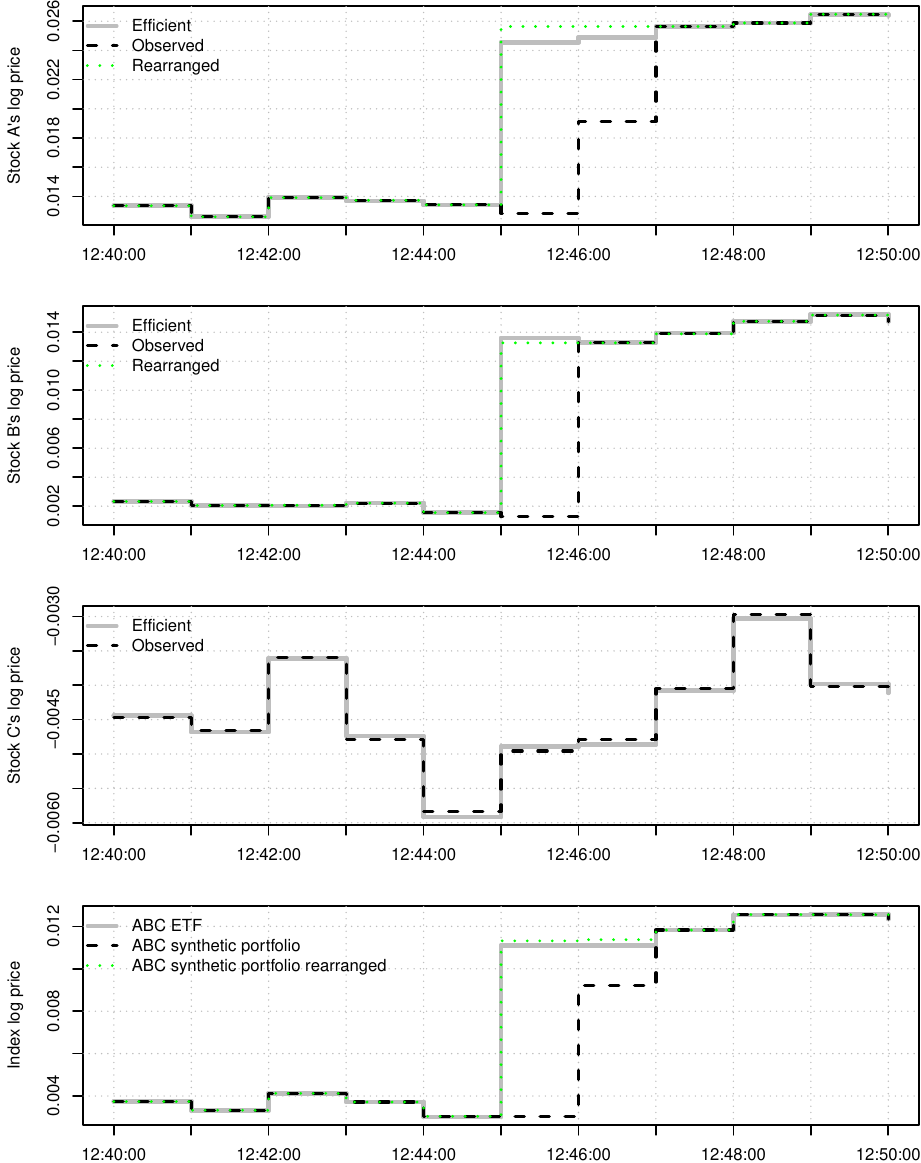}
	\label{fig:Simulate-Multiple-Paths-Rearranged}
	\begin{minipage}{1.0\linewidth}
		\begin{tablenotes}
			\small
			\item {
				\medskip
				Note: 
				This figure 
				shows the efficient (gray), observed (black) and rearranged (green) log prices of the assets. 
				The top panel shows that Stock A jumps gradually and	completes the jump 2 minutes after the ETF in bottom panel, 
				while the second panel shows that stock B's jump is 
				1 minute late and stock C does not jump at all (third panel). 
				The fourth panel shows that the delays cause the implied (inefficient) price of the basket of stocks (the ABC portfolio) 
				to deviate 
				from the efficient ABC ETF price. 
				The best rearrangement approximately recovers the latent, efficient stock price. 
				That is, by shifting the jumps of Stock A and Stock B to earlier periods (top 2 panels), the LP minimizes the distance between the price of the ETF and the synthetic portfolio (lowest panel).
				}
		\end{tablenotes}
	\end{minipage}
\end{figure}

Figure \ref{fig:Simulate-Multiple-Paths-Rearranged} shows the implied prices after this % CN rm: new 
optimal arrangement of jump returns. 
Prices asynchronously incorporate news. 
That is, the observed (black) prices of stock A and stock B deviate from their efficient (gray) values in the first and second panels.  
(There are also small deviations in Stock C's observed prices due to a (relatively smaller) contamination of the continuous component.)
The fourth panel shows that the delays cause the implied (inefficient) price of the basket of stocks (the ABC portfolio) 
to deviate from the efficient ABC ETF price. 
The best rearrangement combines two small jumps of stock A and shifts the jumps of stock B one period backward, aligning the jumps in time. 
The rearranged price paths (green) are now much closer to the efficient price paths (black) in the 1st, 2nd and 4th panel.

% \clearpage

%%%%%%%%%%%%%%%%%%%%%%%%%%%%%%%%%%%%%%%%%%%%%%%%%%%%%%%%%%%%%%%%%%%%%%%%%%%%%%%%%%%%%%%%%%%%%%%%%%%%%%%%%%%%%%%%%%%%%%%%%

\section{Empirics}
\label{secEmpirics}

We apply our methods to 
investigate the reactions of stock prices 
in event windows around ETF jumps. 
We illustrate that synchronizing mistimed stock returns increases the Sharpe ratio of a portfolio allocation strategy. 

 \subsection{Data: The Dow and the DIA ETF}

The NYSE Trade and Quote (TAQ) database provides equity trade data with millisecond precision timestamps.  
The basket instrument is the SPDR Dow Jones Industrial Average ETF (DIA). 
We compare this ETF to the price of a synthetic index of Dow 30 stock prices, as in 
\citet[][]{bollerslev2008risk}.\footnote{
	The constituency of the Dow 30 index is dynamic. The 43 unique tickers within our sample period are: 
	AA,   AAPL, AIG,  AXP,  BA,   BAC,  C,    CAT,  CSCO, CVX,  DD,   DIS,  DOW,  DWDP, GE,   GM,   GS,   HD,   HON, HPQ,  IBM,  INTC, JNJ,  JPM,  KFT,  KO,   MCD,  MMM,  MO,   MRK,  MSFT, NKE,  PFE,  PG,   T,    TRV,  UNH,  UTX,	V,    VZ,   WBA,  WMT and  XOM. 
} 

The data cover the period January 3rd, 2007 through April 2nd, 2020, 
which includes several exceptionally turbulent episodes, 
such as
the global housing and credit crisis, the European sovereign debt crisis and the bail-out of Greece, 
the Russian, Greek, Turkish crisis and the 2020 stock market crash. 

We pre-filter the prices as in \citet{barndorff2009realized}. 
We also remove banking holidays, half-trading days, any day where there is more than a two-hour gap 
	between consecutive trades and periods of malfunctioning such as the 2010 flash crash.

\if1\comment
 \blue{Add a table with the 10 largest standardized spreads? Is it a jump day? Is there news on those days? Unfortunately the diagnostic is a flawed measure.} 
 \fi

There are many jumps in the data.
We apply \citet{boudt2011robust}'s modified \citet{lee2007jumps}'s univariate jump test that accounts for intraday periodicity, on one-minute returns (at $\alpha = 0.1\%$) to identify jumps. 
The univariate tests identify 1,710 ETF jumps across 1,163 (jump) days. 
Some days include gradual or multiple ETF jumps. 

There are many asynchronous jumps.
We construct 1,529 [-5,+5]-minute jump-event matrices, around ETF jumps, as in equation \eqref{eqJumpMatrix}. 
When there are multiple ETF jumps within the event window, the event window spans from five minutes before the first ETF jump to five minutes after the last jump. 
% 
% We constrain the RLP in two ways. 
We contrain the RLP in three ways. 
1) Stocks that  already jump with the index, cannot move, because we assume those are efficient. 
2) No stock jumps may be moved earlier than the highly liquid and carefully watched ETF. 
3) No stock jumps may be moved if the ETF jumps within the first 10 minutes or the last 10 minutes of the trading day. 
% 
% After filtering, % Of those 1,529 jump-event matrices, there are $380$ episodes in which at least one stock jumps after the ETF jump, which makes these 380 episodes candidates for rearrangement. 
After imposing these filters, there remain 380 matrices candidate for rearrangement. 
The RLP rearranges stock jumps in {$180$} cases (or 11.8\% of all jump-event matrices).

    \subsection{News and asynchronous jumps: An empirical illustration}
    \label{ssecEmpIllustration}

We investigate sluggish cojumps, \textit{i.e.} 
jumps that occur later than the index (futures) jumps in the DIA. 
% the jumps occur at close but distinct points in time, in the DIA basket instrument and some of its underlying components to illustrate the asynchronous jumps.
Consider the following example: 
on September 18, 2007 the Federal Reserve announced rate cuts, a bold but risky action according to the financial press.%
\footnote{
	The 
	\href{https://www.federalreserve.gov/newsevents/pressreleases/monetary20070918a.htm}{Press release}
	and the related 
	\href{https://www.federalreserve.gov/monetarypolicy/files/FOMC20070918meeting.pdf}{FOMC Meeting Statement}. 
	See also the coverage in The Economist and the Financial Times: \href{https://www.economist.com/node/9833657/print?story_id=9833657}{Bernanke's bounty}, 	\href{https://www.ft.com/content/c91d7af4-6610-11dc-9fbb-0000779fd2ac}{Instant reaction: Response to the Fed},
	\href{https://www.ft.com/content/8171a091-0790-3907-a9b7-511e36029587}{The Short View: Fed decision — it’s all about game theory},
	\href{https://www.ft.com/content/0c92f198-661a-11dc-9fbb-0000779fd2ac}{Overview: US equities and oil surge after rate cuts}, 
	\href{https://www.ft.com/content/116ef120-662f-11dc-9fbb-0000779fd2ac}{Cheering greets Fed announcement}, 
	\href{https://www.ft.com/content/9c0a6592-6b81-11dc-863b-0000779fd2ac}{Fed must weigh inflation against recession}, 
	\href{https://www.ft.com/content/782afd5c-662d-11dc-9fbb-0000779fd2ac}{Bold Fed goes for half-point cut},
	\href{https://www.ft.com/content/d45efd50-6635-11dc-9fbb-0000779fd2ac}{Bank acts boldly to avert recession risk},
	\href{https://www.ft.com/content/b8220180-501b-3e8d-b440-6aaa0d2c2d93}{Fed cut: Pundits speak}, 
	\href{https://amp.ft.com/content/3fb31ed0-6634-11dc-9fbb-0000779fd2ac}{Fed slashes rates}, and
	\href{https://www.ft.com/content/79ba05a4-4386-3bc2-aa44-205ea3f5f0ef}{Feeling ecstatic? Mind the e-Ben-der}
}

\begin{figure}[htb!]
	\centering
	\caption{
		% When the FOMC Statement is released, the spread temporarily expands and then contracts again
		% News and asynchronous jumps. 
		% The spread (between an ETF and a synthetic index) expands and contracts again
		When the FOMC statement is released on Sep. 18, 2007, the synthetic index-ETF spread expands and contracts again
	}
	\centering
			\includegraphics[width=.9\textwidth,angle = -0]{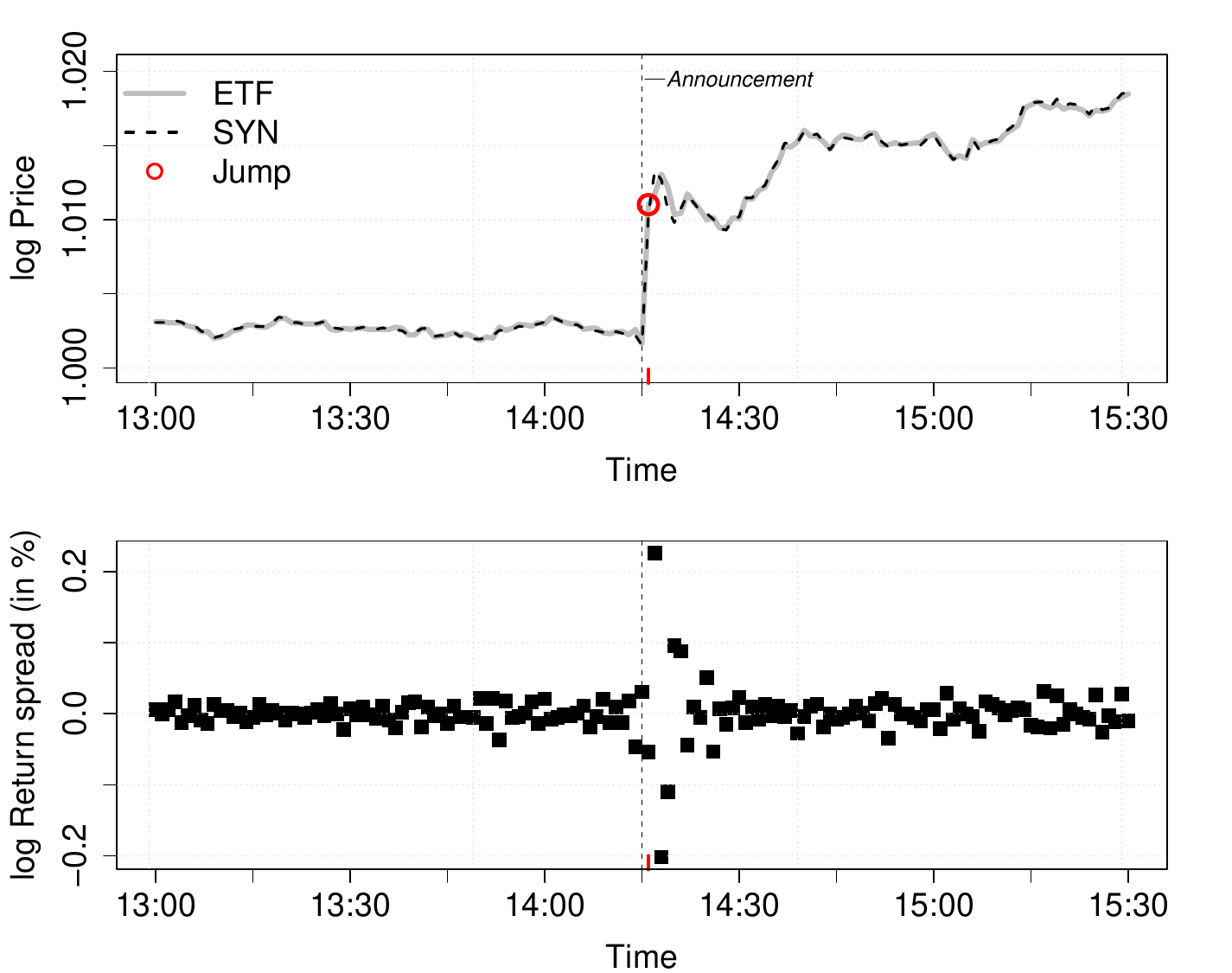}
	
	\label{figMispricingETFDelay}
	\begin{minipage}{1.0\linewidth}
		\begin{tablenotes}
			\small
			\item {
				\medskip
				Note: 
				We plot the one-minute log prices of the DIA ETF (in gray) and a synthetically constructed Dow 30 index (in black), alongside the log return spreads, on the day of the  FOMC Statement of September 18, 2007. 
				The statement gets issued at 2:15 pm and is marked in the top plot by a dashed line. 
				The subsequent ETF jump at 2:16 pm is marked with a red circle and a red vertical tick on the time axis. 
			}
		\end{tablenotes}
	\end{minipage}
\end{figure}

Figure \ref{figMispricingETFDelay} 
shows 
% that some FOMC statements can be difficult to interpret. 
the price paths of the DIA ETF and a synthetic Dow 30 index and the spread between the returns of those two assets  \eqref{eqPriceSpread} on the day of the FOMC Statement -- September 27, 2008. 
\citet{lee2007jumps}'s jump test flags a jump at 2:16 pm, one minute after the release of the statement. 
We mark the ETF jump with a red circle.
The ETF price jumps $0.938\%$ at 14:16, U.S. Eastern Time, one minute after the release. As noted before, if news reached the entire market instantly, was interpreted homogeneously, and trading were continuous, 
jumps in a group of stocks should presumably occur simultaneously with the ETF index jump and 
spreads should be small and random. 
The ETF-synthetic index 
spread temporarily expand, however, following 
the ETF jump, and then contract again. 
That is, markets took time to incorporate the Fed's news into the Dow 30 stock's prices. 

A jump-event matrix \eqref{eqJumpMatrix} % collects 
characterizes 
the asset jumps across an event window from five minutes before to five minutes after the index jump. On Sep. 18, 2007, for example, the event window contains  50 stock jumps, including many gradual jumps, from 30 stocks, of which $27$ match, \textit{i.e.} occur at the same time as, the ETF jump and $23$ lag the ETF jump. 

\begin{figure}[htb!]
	\centering
	\caption{Range of the spreads and matching stocks as a function of the  permitted maximum length of the move}

	\centering
	
	\includegraphics[width=.33\textwidth,angle = -90]{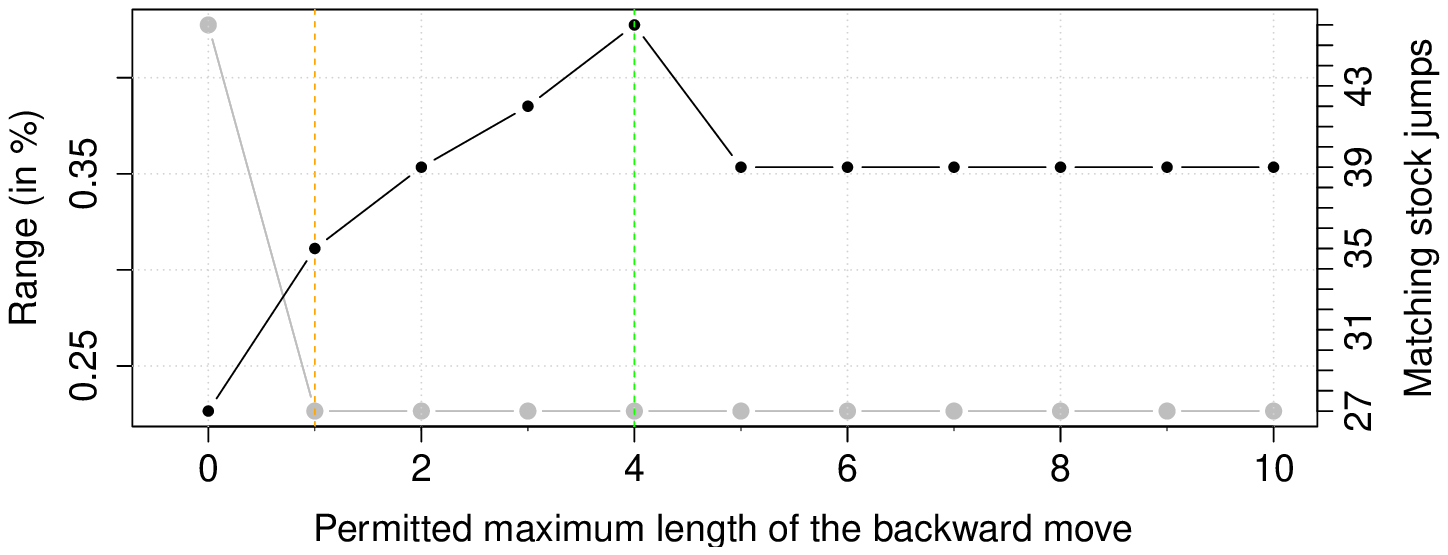} 
	
	\label{figExampleRearrangement}
	\begin{minipage}{1.0\linewidth}
		\begin{tablenotes}
			\small
			\item {
				\medskip
				Note: 
				This figure plots the range
    % , \textit{i.e.} flatness, 
    of the return spreads (left scale, gray) and the number of matching stocks (right scale, black) as a function of the number of the permitted {length of the} move during the Sep. 18, 2007 FOMC statement. 
				The RLP rearranges scattered jumps of the Dow 30 stocks. 
				We allow each jump to move at most 10 minutes backward in time and solve the RLP for each of these possible constraints.
				We mark the rearrangement with minimal range with an {orange} line and the rearrangement with the maximum number of matching stock jumps (for the same minimal range) with a green line. 
			}
		\end{tablenotes}
	\end{minipage}
\end{figure}

Figure \ref{figExampleRearrangement} shows the best rearrangement of the stock jumps for this event.  By ``best rearrangement," we mean the jump arrangement that minimizes the range of the spreads. 
% By ``best rearrangement," we mean the jump arrangement that minimizes the range of the spreads. 
The figure shows the range of the return spreads (left scale, gray) and the number of {matched stocks} (right scale, black) as a function of the permitted length of the move in time for each individual jump, as in equation \eqref{eqPenalty}.
The range declines as we permit {larger moves in time} 
and 
it is the criterion that determines the chosen number of 
backward moves 
for jumps. 
Permitting 
each jump to move one minute backward
minimizes the range (orange line); the range drops from 0.427\% to 0.227\%.
Permitting a maximum backward repositioning length of 4 periods or 4 minutes 
(green line) has the same minimal range for a larger number of matching stocks.
This rearrangement matches $19$ out of $23$ scattered stock jumps with the ETF jump, approximately recovering the common jump in the stocks.

Recovering the common jump is likely to improve estimates of the daily realized covariance matrix. 
The realized covariance matrix is defined as \citep{barndorff2004econometric}: 
\begin{eqnarray}
	\label{eqRC}
	% We already use t for diffusion process
	{{C}}_d = \sum_{i=1}%
	%^n
	% Salome
	^{\floor{T/\Delta_n}}
	(\Delta_i^n {Y})
	(\Delta_i^n {Y})^\top
	\end{eqnarray}
in which, 
${Y}_{i\Delta_n} = (Y_{1,\Delta_n}, ..., Y_{p,\Delta_n})^\top$ 
is the observed log price process \eqref{eqObservedPrice}
 of the $p$ stocks sampled on a regular time grid $\{i\Delta_n: 0 \leq i \leq \floor{T/\Delta_n} \}$ over one day $T = 1$, 
the $i$th return of ${Y}_{i\Delta_n}$ is 
$\Delta_i^n {Y} = {Y}_{i\Delta_n} - {Y}_{(i-1)\Delta_n}$. 
The standard realized covariance matrix \eqref{eqRC} plugs in raw returns. 
Other estimators, like the multivariate realized kernel in 
\citet[][]{barndorff2011multivariate} or a Cholesky factorization in \citet{boudt2017positive},  protect against  mild market microstructure noise 
and the \citet{epps1979comovements} effect, that is, the downward bias in 
covariance estimates due to asynchronous trading. 
Using rearranged returns in \eqref{eqRC} also protects against  asynchronous jumps and the underestimation of jump dependence. 

\begin{figure}[htb!]
	\centering
	\caption{The realized covariance is different in the rearranged returns on the day of the FOMC statement}

	\centering

		\subfloat[Realized covariances of HPQ: Sep. 18, 2007]{
		\includegraphics[width=.50
		\textwidth,angle = 0]{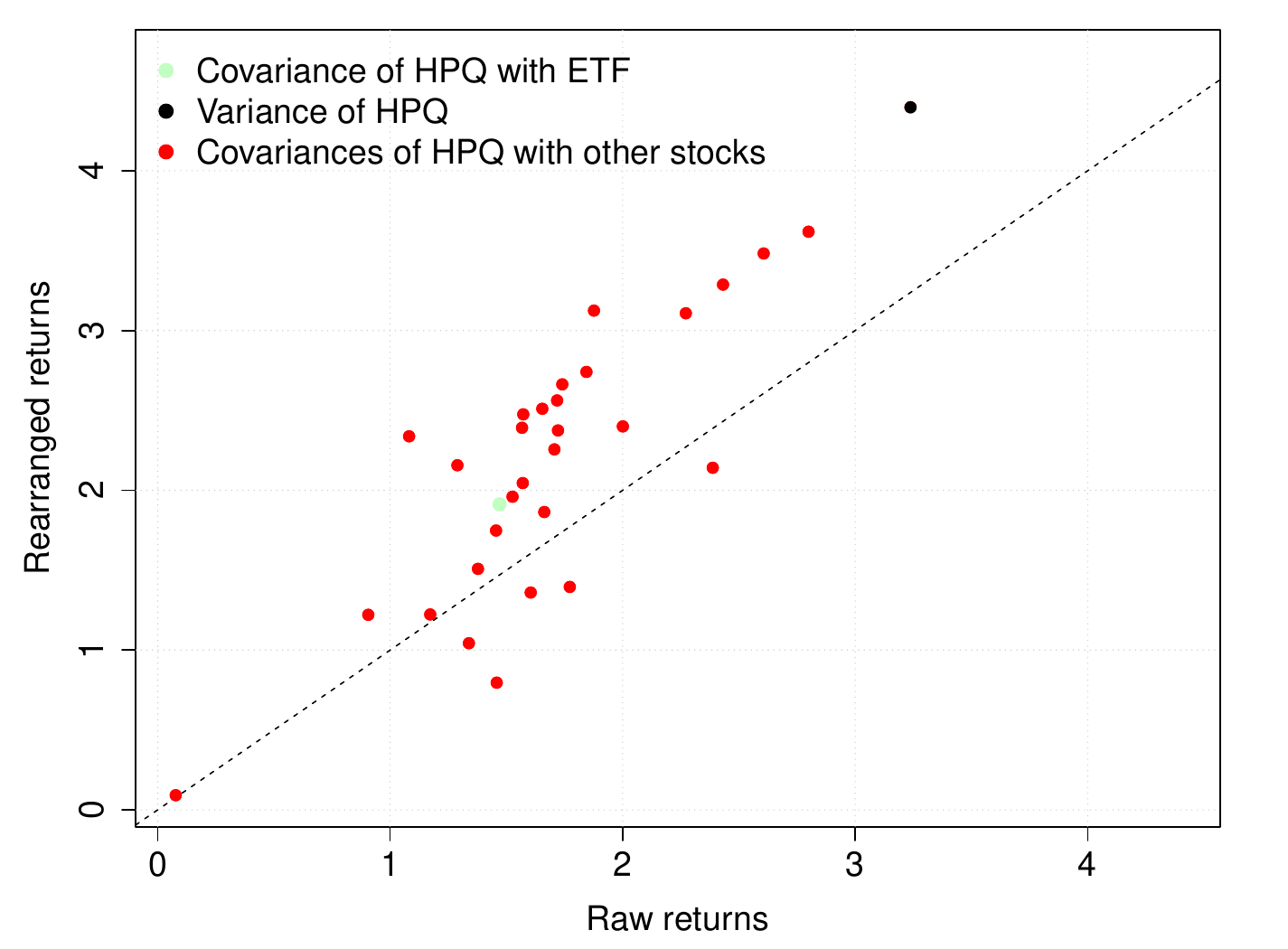} 
	}
	\subfloat[Realized covariances of PFE: Sep. 18, 2007]{
		\includegraphics[width=.50
		\textwidth,angle = 0]{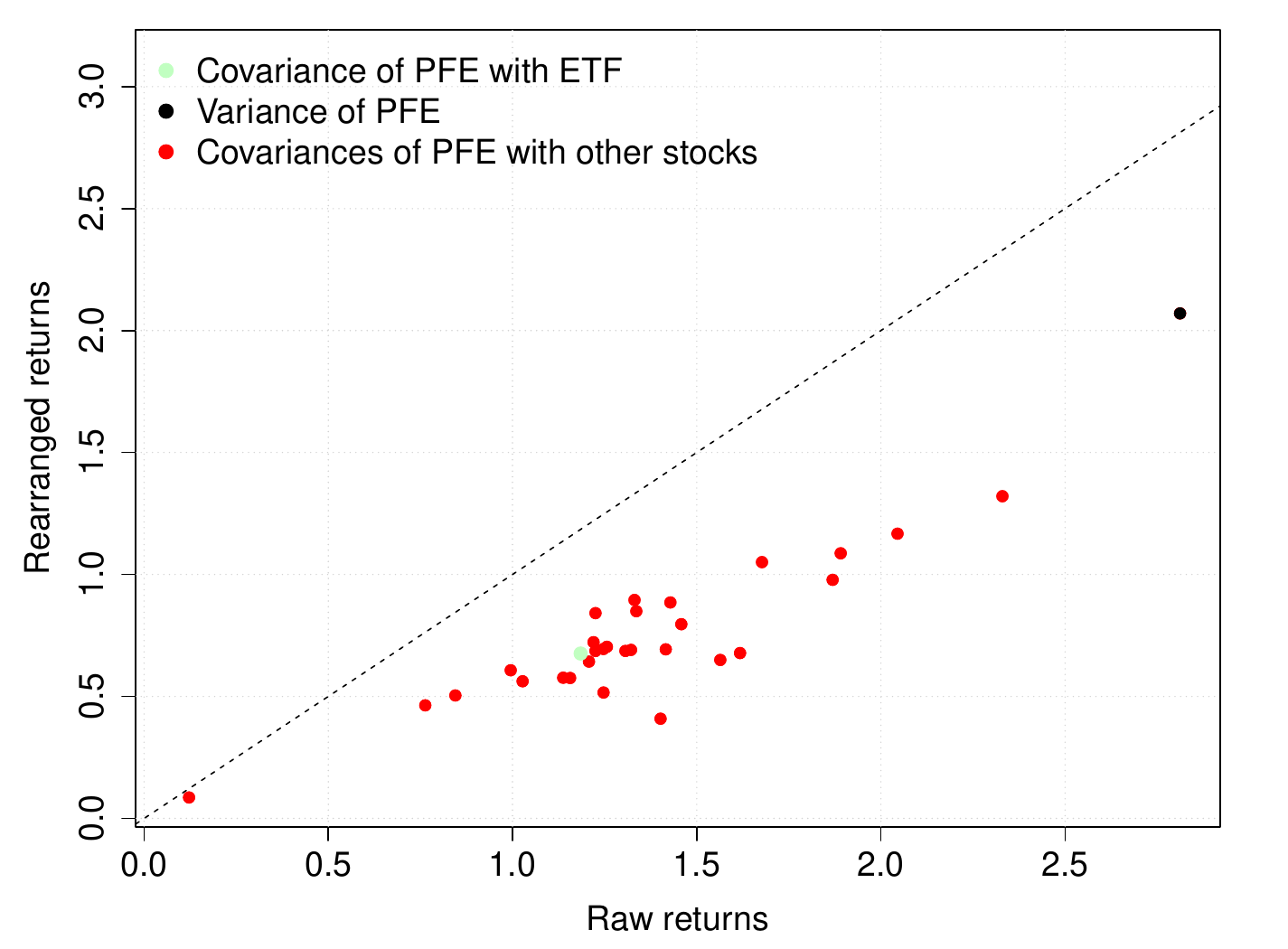} 
	}

	\label{figExampleRearrangementBetas}
	\begin{minipage}{1.0\linewidth}
		\begin{tablenotes}
			\small
			\item {
				\medskip
				Note: 
				This figure shows  
				two scatter plots consisting of 
				the elements in the 
				realized covariance matrix for 
				two 
				stocks 
				(HPQ and PFE), relative to DIA and the other Dow 30 stocks, 
				on the day of the FOMC statement. 
				The x-axis shows the realized covariances with raw returns and the y-axis the realized covariances with rearranged returns. 
			}
		\end{tablenotes}
	\end{minipage}
\end{figure}

Figure \ref{figExampleRearrangementBetas} allows us to visually compare covariance estimates made with raw \textit{vs.} rearranged returns. 
It suggests that % synchronizing stock jumps 
optimally changing the time labels of one of two observations (out of the 390 one-minute returns required to estimate the realized covariance) 
changes the estimated covariance structure between the stocks and the ETF returns on the day of the FOMC statement. 
 For example,  the fact that the statistics in  the left panel generally lay above the 45-degree line shows that rearranging the jumps of HPQ, which jumps gradually,  increases its variance and most covariance with other stocks. 
 The right panel shows that rearranging the jumps of PFE, which jumps with an overreaction, reduces its own variance and all the covariances with PFE returns. 
 % The linear program typically only changes the time labels of only one or two observations (out of the 390 one-minute returns required to estimate the realized covariance). 
 
% To test the usefulness of using rearranged returns, compared to using raw returns, we use the realized covariance matrices in building minimum-variance portfolios \eqref{ssecMarkowitz}. 
% If there's an additional application. 

A practical question is whether one should use either raw returns or  rearranged returns. 
Our recommendation is to always use the rearranged returns. 
These synchronized returns are more precise in a high-frequency analysis of market reactions around jumps. 

\subsection{Minimum-variance portfolios}
\label{ssecMarkowitz}

% To evaluate the economic value of using rearranged returns, 
We show the usefulness of using rearranged returns, compared to using raw returns, in the context of portfolio allocation, which shows that high-frequency rearranged jump returns affect the performance of low-frequency decisions, like building a daily-rebalanced minimum-variance portfolio. 

Stock returns determine the weights that minimize the portfolio variance. 
The optimal weights for a particular day minimize the portfolio variance, subject to the constraints that they deliver a given expected return and that they sum to one \citep[][]{markowitz52}: 
\begin{eqnarray}
	\label{eqMarkowitz}
	\underset{w_d}{\text{Minimize}} \, \, \, 
	\sigma_{p,d}^2 = {w}_d' 
	C_d
	{w}_d, 
	\, 
	\text{ s.t. } 
	{w}_d' {\mu} = \mu_{p,d} \text{ and } {w}_d' {1}, 
	\end{eqnarray}
in which  
{$\sigma_{p,d}^2$}
is the daily variance of the portfolio return, 
${w}_d = (w_{1,d}, ..., w_{p,d})^\top$ is the daily $p$-dimensional weight vector, $C_d$  is the daily realized covariance matrix and $\mu_{p,d}$ is the {target} portfolio return. We make a grid of 100 target returns that range from the lowest average stock return to the highest one. 
% As in the previous section, we compare  two ways to calculate the realized  covariance: with raw returns and with rearranged returns. 
To test our synchronization procedure, compare a simulated portfolio’s performance using raw vs. rearranged returns to compute the realized covariance matrix. % $C_d$.
% The portfolio weights are functions of the estimated realized covariance matrix.

We optimize weights \eqref{eqMarkowitz} on rearrangement days, \textit{i.e.} the 184 ETF jump days for which we rearrange jumps  and rebalance the portfolio the next day.
% We rebalance the portfolio with these optimal weights the day after such a rearrangement day. 
We keep those weights until the next rearrangement day. 
In those cases when the Dow constituency changes in between rearrangement days,  we reset to an equally weighted portfolio until the next rearrangement day.

Table \ref{tabMarkowitz} reports the portfolios' closing value, 
standard deviation and {modified} Sharpe ratio at the 5\% level and the $p$-value of their difference \citep{ardia2015testing} for the full sample and for each year. 
% We observe superior out-of-sample performance of using rearranged returns, as opposed to raw returns in a trading strategy. 
For {12 of 14} years and over the whole sample, the rearranged-return portfolio statistics are superior to the raw-returns portfolio statistics in terms of the modified Sharpe ratio. Over the full sample, the modified Sharpe ratios are significantly different and the rearranged-return portfolio delivers an additional 5\% performance.

\begin{table}[htb!] % htb!
	% #4_Null_Simulation_seq_OtherMethods
	\centering
	\caption{Performance of minimum-variance portfolios}
	\label{tabMarkowitz}
	\begin{adjustbox}{max width=\textwidth}
		\begin{tabular}{r | rrr | rr | rr | rrr | rrr }
			&\multicolumn{1}{c}{Days}
			% &\multicolumn{1}{c}{CJ-Days}
			&\multicolumn{1}{c}{\#CJ}
			% &\multicolumn{1}{c}{CJ-R-Days}
			&\multicolumn{1}{c}{\#CJ-R}
			&\multicolumn{2}{c}{Closing value}
			% 
			% &\multicolumn{2}{c}{Means}
			&\multicolumn{2}{c}{SD}
			% &\multicolumn{3}{c}{Sharpe}
			&\multicolumn{3}{c}{mSharpe}
			% &\multicolumn{2}{c}{Min.}
			% &\multicolumn{2}{c}{Max.}
			\\
			&&&
			% &\multicolumn{1}{r}{Eqw}
			% &\multicolumn{1}{r}{Raw}
			% &\multicolumn{1}{r}{DIA}
			&\multicolumn{1}{r}{Raw}
			&\multicolumn{1}{r}{Rearr.}
			% &\multicolumn{1}{r}{Eqw}
			%			&\multicolumn{1}{r}{Raw}
			%			&\multicolumn{1}{r}{Rearr.}
			%			% 
			%			&\multicolumn{1}{r}{Raw}
			%			&\multicolumn{1}{r}{Rearr.}
			% &\multicolumn{1}{r}{DIA}
			&\multicolumn{1}{r}{Raw}
			&\multicolumn{1}{r}{Rearr.}
			% 
			% &\multicolumn{1}{r}{DIA}
			% &\multicolumn{1}{r}{DIA}
			&\multicolumn{1}{r}{Raw}			
			&\multicolumn{1}{r}{Rearr.}
			& \multicolumn{1}{r}{$p$-value}
			% &\multicolumn{1}{r}{Raw}			
			% &\multicolumn{1}{r}{Rearr.}
			% & \multicolumn{1}{r}{$p$-value}
			\\
			\hline
			% \multicolumn{1}{l}{Nominal level (in\%)} & 5.000 \\ 
			% \multicolumn{1}{l}{Dimension ($d$)} & 300\phantom{.000} \\ 
			% 
			Full period 
			& 3,264 & 1,163 &  {184}
			% & 1.3500 
			& 1.8500 & \grb 1.9000
			% & 0.026 & 0.028
			% & 0.741 & 0.741
			% & 0.035& \grb 0.037 \\ 
			% & 0.9140   
			&  0.7410 & 0.7410 
			% & & & 
			& 0.0159 & \grbb 0.0178	&  0.0879
			\\
			%			Subsample & & & 184 
			%			& -- & -- 
			%			% mean 0.0432 0.0456 
			%			& 0.6740 & 0.6710
			%			\\ 
			%%%%%%%%%%%%%%%%%%%%%%%%%%%%%%%%%%%%%%%%%%%%%%%%%%%%%%%%%
			Year-by-year \\
			2007 
			& 246 & 57 & 13 
			 % & 0.9365 
			%  &  1.0202  & \grb 1.0301  % before
			&  1.0191   & \grb 1.0294
			% & -0.0259 & 0.0082 & 0.0123
			% 0.7560  & 0.5583 & 0.5645
			% & -0.0343 
			% &  0.0147 & \grb 0.0217 % sr before
			% means
			% &  0.008 & 0.012	
			% sd
			& \grb 0.5592 & 0.5652
			% & --    
			&  0.0357 & \grb 0.0430
			& 0.2520
			\\
			2008 
			& 247 & 55 & 6
			% & 0.6624 
			% & 0.8094 
			&  \grb 0.9900 & 0.9880
			% & 1.9524 
			% & 1.5156 & \grb 1.5043
			& 1.5156 & \grb 1.5043
			& 0.0377 & \grb 0.0378
			& 0.9970
			\\ 
			2009  
			& 243 & 68 & 11
			% & 0.9943 
			% &   1.0638 &  \grb 1.0703
			% & -0.0015 
			% &  0.0286 & \grb 0.0314
			&  1.0607 & \grb 1.0684 
			& \grb 0.9220 & 0.9254
			& -0.0127 & \grb -0.0110
			& 0.2700
			\\
			2010 
			& 247 & 65 & 5
			% & \grbb 1.0483 
			% & 1.0207 & \grb 1.0378
			& 1.0213 & \grb 1.0384
			% & \grbb 0.0245 
			% & 0.0133 & \grb 0.0242
			& \grb 0.6292 & 0.6324
			& -0.0373 & \grbb -0.0295
			& 0.0919
			\\
			2011 
			& 247 & 72 & 6
			% & 0.9343 
			& 1.1505 & \grb 1.1537
			% & -0.0282 
			% &  0.0965 & \grb 0.0999
			& 0.6268 & \grb 0.6219
			&  0.0628 &  \grb 0.0644
			& 0.5500
			\\ 
			2012 
			& 245 & 107 & 8
			% & \grbb 1.0719 
			% & 0.9439 & \grb 0.9490
			&  0.9439 & \grb 0.9493
			% & \grbb 0.0470 
			% & -0.0456 & \grb -0.0413
			& \grb 0.5045 & 0.5056
			& -0.0634 & \grb -0.0609
			& 0.2950
			\\ 
			2013 
			& 245 & 90 & 13
			% & \grbb 1.1056 
			% & \grb 1.1042 & 1.0954
			% & 1.104 & 1.095
			& \grb 1.1049 & 1.0965
			% & \grb 0.0878 
			% & \grb  0.0879 & 0.0805
			& 0.4865  & \grb 0.4864
			& \grb  -0.0034 & -0.0101
			& 0.1130
			\\ 
			2014 
			& 247 & 78 & 15
			% & 1.0386 
			% & \grb 1.2593 &  1.2499
			% now better than before?
			% Raw = 
			% & 1.259 & 1.250
			& \grb 1.2607 & 1.2510 
			% & 0.0299 
			% & \grb 0.2201 &  0.2147
			&  0.4790  &  \grb 0.4730
			&  \grb 0.2310 & 0.2260
			& 0.5180
			\\ 
			2015 
			& 247 & 72 & 9
			% & \grbb 1.0416 
			% & 1.0135 & \grb 1.0210
			& 1.0135 & \grb 1.0211
			& 0.6918  & \grb 0.6843
			& -0.0061 & \grb -0.0006
			& 0.4530
			% & \grbb 0.0219 
			% & 0.0079 & \grb 0.0125
			\\ 
			2016 
			& 249 & 101 & 14
			% & 1.1168 
			% & 1.1470 & \grb 1.1475
			& 1.1480 & \grb 1.1482
			% & 0.0849 
			% &  0.1158 & \grb 0.1163
			& 0.5113   & \grb 0.5109
			& 0.0118 & \grb 0.0119
			& 0.9860
			\\ 
			2017 
			& 247 & 133 & 23
			% & 1.0813 
			% & 1.1229 & \grb 1.1248
			% & 1.123 & 1.125
			& 1.1220 & \grb 1.1234
			& \grb 0.2977  & 0.2986
			& 0.0692 & \grb 0.0709
			& 0.6880
			\\
			2018 
			& 245 & 93 & 14
			% & 0.8677 
			% &  0.9240 & \grb 0.9323
			% & 0.924 & 0.932
			& 0.9231 & \grb 0.9310 
			% & -0.0566 
			% &  -0.0385 & \grb -0.0335
			& \grb 0.8102 & 0.8278
			& 0.0254 & \grb 0.0302
			& 0.4400
			\\ 
			2019 & 246 & 140 & 42
			% & 1.0559 
			% &   1.0829 & \grb 1.0924
			& 1.0817 & \grb 1.0921 
			% & 0.0422 
			% &  0.0667 & \grb 0.0737
			& \grb 0.5076  & 0.5123
			& -0.0028 &  \grb 0.0040
			& 0.2380
			\\ 
			2020 
			& 63 & 32 & 5
			% & 1.0210 
			% &  1.0129 & \grb 1.0135
			& 1.0135 & \grb 1.0140
			% & 0.0204 
			% &  0.0136 & \grb 0.0142
			& \grb 1.5388  & 1.5407
			& -0.0021 & \grb -0.0017
			& 0.8820
			\\
			\hline
			\hline
			
	\end{tabular}}
\end{adjustbox}	
	\begin{minipage}{1.0\linewidth}
	\begin{tablenotes}
		\small
		\item {
			\medskip
			Note: 
			This table reports the performance of the daily returns of a minimum-variance portfolio optimization with raw \textit{vs.} rearranged returns. 
			From left to right, the columns 
			report the number of days (Days), the number of ETF jumps or cojump days (\#CJ) and
			the number of cojump days on which the rearrangement moves stock jumps (\#CJ-R). 
			``Closing value" is the end-of-sample portfolio value if the initial value was  \$1 using cumulative daily returns. 
			``SD" and ``mSharpe" are the 
			 standard deviations and {modified} Sharpe ratios 
			 and the $p$-value of their difference \citep{ardia2015testing}
			of the portfolio over the sample. 
			Gray cells denote the higher of each performance statistic 
			two Sharpe ratios. 
			{Dark gray cells denote the significant differences of the two Sharpe ratios at a significance level of $10\%$. 
				To calculate the modified Sharpe ratio, we use excess returns \textit{vs.} the DIA ETF. 
				To perform the inference of the difference in Sharpe ratios, we rely on the studentized circular bootstrap approach, with 1000 bootstrap replications and a block length of 5 in the circular bootstrap.}
		}
	\end{tablenotes}
\end{minipage}
\end{table}

\if1\comment
\blue{The added value of rearranging returns is small.}
\fi

% \clearpage

%%%%%%%%%%%%%%%%%%%%%%%%%%%%%%%%%%%%%%%%%%%%%%%%%%%%%%%%%%%%%%%%%%%%%%%%%%%%%%%%%%%%%%%%%%%%%%%%%%%%%%%%%%%%%%%%%%%%%%%%%

\section{Concluding remarks}
\label{secConclusion}

Stock prices often react sluggishly to news, producing gradual jumps and jump delays. 
% In a panel of high-frequency intraday stock returns, 
% the jumps 
% occur 
% at close but distinct points in time. 
% We introduce a new object, 
% the jump-event matrix, to analyze the asynchronous jumps in the ETF and the corresponding cojumps in the underlying stocks. 
The spread between the ETF price and the price of a synthetically constructed index measures the collective misalignment of noisy stock prices with their respective equilibrium levels. 
We introduce tools to synchronize the scattered jumps in a 
jump-event matrix 
and better approximate the efficient common jump.  
The rearrangement is currently very practical for problems with up to 30 stocks. We are working on a block rearrangement to allow for larger dimensions and, for example, synchronize stock jumps in the S\&P500 index.

Estimating realized covariance matrices with these synchronized stock returns, as opposed to using raw returns, 
improves
out-of-sample 
portfolio performance. 
Recovering the common jump on a fine sampling grid  is likely to improve other asset allocation and risk management decisions, like 
estimating 
the jump size distribution \citep[see \textit{e.g.},][]{boudt2011outlyingness}, 
estimating 
 jump dependence 
 \citep*[see \textit{e.g.},][]{li2017mixed,li2017jump,li2019rank}
or
forecasting 
realized measures 
(see \textit{e.g.}, \citeauthor{andersen2007roughing}, \citeyear{andersen2007roughing}; 
\citeauthor{bollerslev2020realized}  \citeyear{bollerslev2020realized}; 
\citeauthor{bollerslev2022realized}, \citeyear{bollerslev2022realized}). 
A thorough analysis must, however, await future work. 

% \clearpage

\bibliographystyle{chicago}
\bibliography{Bibliography}

%%%%%%%%%%%%%%%%%%%%%%%%%%%%%%%%%%%%%%%%%%%%%%%%%%%%%%%%%%%%%%%%%%%%%%%%%%%%%%%%%%%%%%%%%%%%%%%%%%%%%%%%%%%%%%%%%%%%%%%%%

\appendix

\section{Simulating sluggish news reactions}
\label{secSimExample}

We show how to generate a sample path from the new DGP, in which the discontinuous component is spread across several time intervals. 
We first simulate second-by-second ($\Delta_n = 1/23,401$) efficient log prices \eqref{eqEfficientPrice} for 1 stock ($p = 1$), $X_{i\Delta_n}$,  across one trading day ($T = 1$) from a jump-diffusion process. 

\subsection{The continuous component}

The continuous component, $X^c_{i\Delta_n}$, has a  zero drift and constant variance $\sigma^2_t = 0.039$, corresponding to an annualized return volatility of 20\%. 
We add \textit{i.i.d.} microstructure noise $u$, with $E[u] = 0$ and $E[u^2] = \omega^2$, in which we select $\omega^2$ by fixing the noise ratio to $\gamma = (n \omega^2 / \int^1_0 \sigma_s^2 ds)^{\nicefrac{1}{2}} = 0.5$, to contaminate the continuous part of the efficient price \citep[as in][]{christensen2014fact}. 

\subsection{The discontinuous component}
% We draw the discontinuous component $X^d_{i\Delta_n}$,  from a compound Poisson process: 
Econometricians typically assume a compound Poisson for the efficient jump process: 
\begin{align}
	\label{eqSimPoisson}
	X^d_{i\Delta_n} \equiv 
	\sum^{N^J}_{j = 1} {I}_{\{U_{j} \cdot T  \leq i\Delta_n\}} 
	\Delta X_{j},
\end{align}
in which $I(\cdot)$ is an indicator function, 
$N^J$ is the number of jumps that occur during a day, 
$U_j$ are the random arrival times of the jumps 
and the sequence of normally distributed random variables, $\Delta X_{1}, ...,  \Delta X_{N^J}$,
are normally distributed jump sizes.

The news that generates the jump in the efficient price is not immediately impounded in the stock's observed price. 
To simulate such a sluggish news reaction, we draw a contaminated jump process,  $Y^d_{i\Delta_n}$, 
which includes a step function that spreads each individual efficient jump size,  $\Delta X_{j}$,  within the compound Poisson process \eqref{eqSimPoisson}, across several time intervals: 
\begin{align}
	\label{eqSimObsJump}
	Y^d_{i\Delta_n} 
	\equiv 
	\sum^{N^J}_{j = 1} {I}_{\{U_{j} \cdot T  \leq i\Delta_n\}} 
	% \left( 
	\underbrace{
	\sum^{N_j^D}_{d = 0} {I}_{\{W_{j,d} \cdot T  \leq i\Delta_n\}} \Delta L_{j,d} 
	}_{\text{Progress to efficiency}}
	\, 
	\Delta X_{j}.
	% \right),
\end{align}
in which $N_j^D$ governs the total number of steps % in the step function, 
within each jump's delay process, 
$W_{j,0}, ..., W_{j,N_j^{D}}$ are the step arrival times and $\Delta L_{j,0}, ..., \Delta L_{j,N_j^{D}}$ are increments in the step sizes, 
which add chunks of the efficient jump size, $\Delta X_j$, to the observed jump process, $Y^d$. 
The step function captures the progress to effiency;
%  the speed with which observed prices incorporate new information: 
it rises from $0$ to $1$ as information about the efficient jump is fully incorporated in the stock's observed price.

\subsubsection{Step widths}

\sloppy
% Overflowing equation in the margin. 
Suppose an efficient price jumps 
at the random arrival time, $U_j$, with a jump size, $\Delta X_j$. 
To simulate the step function within 
for this particular jump, 
we first draw the number of steps,  $N^{D}_j$, from a binomial distribution: 
\begin{eqnarray*}
N^{D}_j \sim \text{Bin} (\text{Number of trials} = 5, \text{ Success probability} = 0.4), 
\, \text{for} \, j \, \text{fixed}.
\end{eqnarray*}

Each of the steps has a random width. 
An exponential distribution governs the waiting times between the steps:  
\begin{eqnarray*}
w_{j,d} \sim \lceil \text{Exp} [ \text{Rate} = 1 / (15 \, N^{D}_j)] \rceil, 
\, \text{for} \, j \, \text{fixed} \, \text{and} \, d = 1, ..., N^{D}_j,
\end{eqnarray*}
in which $\lceil \cdot \rceil$ is the ceiling operator. 
The step arrival times, $W_{j,d}$, in the indicator function of the compound information process are the cumulative sums of the waiting times starting at the arrival time of the efficient jump $U_j$: \, 
\begin{eqnarray*}
W_{j,0} \cdot T  		\equiv& U_j \cdot T &,\\
W_{j,1} \cdot T   	 \equiv& U_j \cdot T &+ w_{j,1}, \\
W_{j,1} \cdot T   	 \equiv& U_j \cdot T &+ w_{j,1} + w_{j,2},\\
\vdots \\ 
% W_{j,N^D_j} \cdot T \equiv& U_j \cdot T &+ w_{j,1} + ..., w_{j,N^D_j}.
% \equiv U_j \cdot T+ D_j
W_{j,N^D_j} \cdot T \equiv& U_j \cdot T &+ D_j.
\end{eqnarray*}
The starting point of the step function, $W_{j,0}$, is the arrival time of the efficient jump. 
The end point,  $W_{j,N^D_j}$, or the moment when 
the 
information has been fully impounded, is the starting point plus the sum of the waiting times. 
The total delay with which the efficient stock jump is impounded in the observed price is the sum of the waiting times: $D_j = \sum_{d = 1}^{N^{D}_j} w_{j,d}.$ 
If it takes more steps to incorporate the efficient jump, the total delay will be longer. 

\subsubsection{Step sizes}

The steps also have random sizes that correspond to accumulated information impounded in the observed price. 
To extract the sizes, $L_{j,d}$, of each step $d$, with $d = 0, ..., N^{D}_j$,
we sample realizations from a Brownian bridge according to the step arrival times, $W_{j,0}, ..., W_{j,N_j^D}$. 

\subsubsection*{The Brownian bridge}

The Brownian bridge 
models 
the latent news impoundment  process. 
When news is released, the market processes and accumulates information, including under- and overreactions. 
Eventually, 
the price fully and correctly impounds the information. 

The Brownian bridge, $\Lambda_{j,t}$, is a continuous-time, stochastic process defined as: 
\begin{align*}
	\Lambda_{j,t} 
	= 
	B_{j,t} - \frac{t}{U_j \cdot T + D_j} 
	\left(1 - 
	% W_{j,T}
	B_{j,U_j  \cdot T + D_j)}
	\right), 
	\qquad t \in [U_j \cdot T, U_j \cdot T + D_j]
\end{align*}
in which 
$B_{j,t}$ is a standard, univariate Wiener process, with $B_{j,0} = 0$. 
A standard Wiener process is tied down to the origin, but the other points are not restricted. 
The Brownian bridge is pinned at both ends
 interval, at $t = U_j \cdot T$ and $t = U_j \cdot T + D_j$. 
 Just as pylons support a literal bridge, the pylons in the Brownian bridge make sure that the process evolves from the first pylon $\Lambda_{j,U_j \cdot T} = 0$ to the second pylon $\Lambda_{j,U_j \cdot T + D_j} = 1$. 
 
\subsubsection*{Sampling step sizes from the Brownian bridge}

The variable governed by the Brownian bridge gradually moves from 0 to 1 -- non-monotonically -- but we observe its values at discrete intervals, 
because 
investors impound 
chunks of new information in the observed price at 
each interval. 
We sample the step sizes from the Brownian bridge process $\Lambda_{j,t}$ at discrete points, $W_{j,0}, W_{j,1}, ..., W_{j,N^D_j}$. 

The Brownian bridge and the waiting times exist in continuously observed prices and are functions of the data generating features that delay jumps. They have nothing to do with the data frequency used by the econometrician. 

\subsubsection{Example of a step function}

\begin{figure}[htb!]
	\caption{A step function captures the speed with which  
		% models how 
		% observed prices incorporate new information 
		observed prices incorporate new information
	}
	\centering
	
	\includegraphics[width=.9\textwidth,angle = -0]{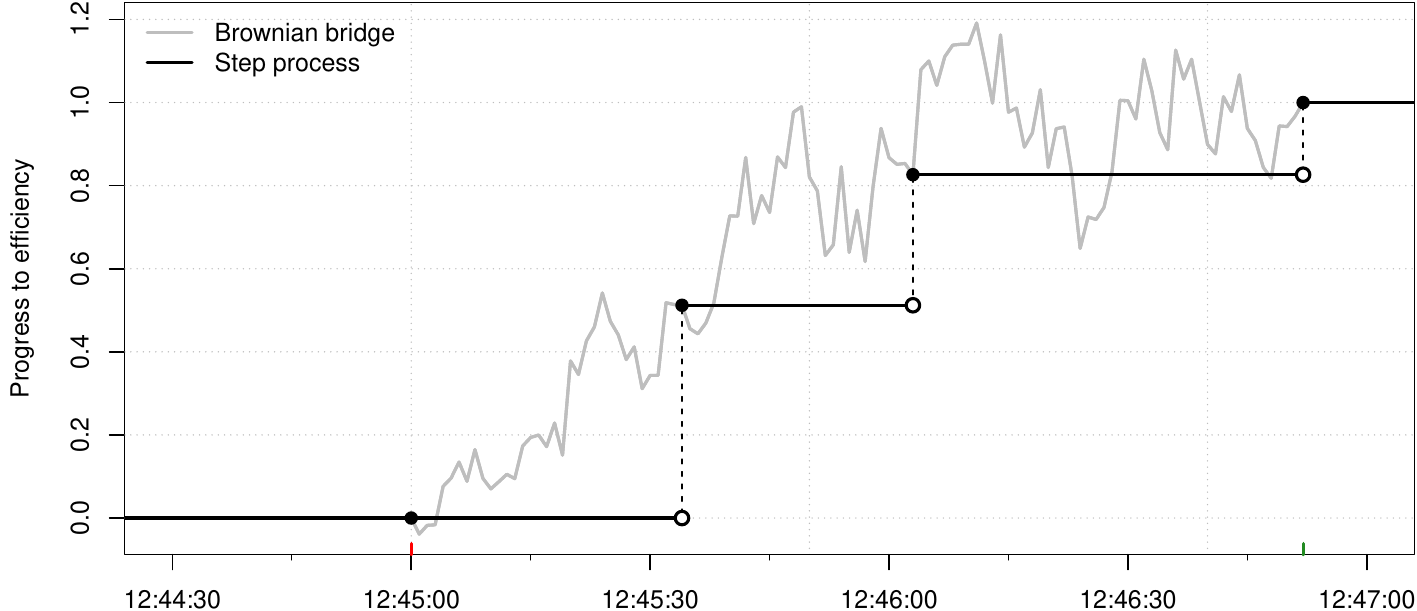}
	\label{fig:Stochastic-Step-Function}
		\begin{minipage}{1.0\linewidth}
			\begin{tablenotes}
				\small
				\item {
					\medskip
					Note: 
					This figure plots a sample path of the step function we use to contaminate an efficient jump.  
					The efficient stock 
					price jumps 
					at $i\Delta_n = 11,701$ or
					12:45:00.
					The waiting times between each of three steps are equal to $w_{1,1} = 34, 
					w_{1,2} = 29, w_{1,3}= 49$ seconds. 
					The sampled sizes are equal to % $L_{1,0} = 
					$0.000$, $0.512$, $0.826$, and $1.000$ and 
					the increments are equal to $\Delta L_{1,0} = 0.000$, $\Delta L_{1,1} = 0.512$, $\Delta L_{1,2} = 0.314$, and $\Delta L_{1,3} = 
					0.174$. 
				}
			\end{tablenotes}
		\end{minipage}
\end{figure}

Figure \ref{fig:Stochastic-Step-Function} plots an example of such a step function for one delayed jump. The efficient stock price jumps 
at $i\Delta_n = 11,701$ or 12:45:00.
The waiting times between each of three steps are equal to $w_{1,1} = 
34, 
w_{1,2} = 29, w_{1,3}= 49$ seconds. 
In the following $D_j = 112$ seconds, 
the Brownian bridge (in black) process evolves from $0$ to $1$.
 If ${\Lambda}_{j,i\Delta_n} < 1$, the observed price jump underreacts and does not (completely) incorporate the new information. 
If ${\Lambda}_{j,i\Delta_n} > 1$, the observed price overreacts to the jump. 
At the end of the interval, when ${\Lambda}_{j,i\Delta_n} = {1}$, the observed stock jump equals
the efficient stock jump. 

We do not observe this learning process in continuous time. Rather, we sample the step sizes at the waiting times, leading to the step function (in blue). 
The waiting times relate to the step widths and are equal to 
$W_{1,0} \cdot T = 11,701 + 0$, $W_{1,1} \cdot T = 11,701 + 34$, $W_{1,2} \cdot T = 11,701 + 63$ and $W_{1,3} \cdot T = 11,701 + 112$. 
The sampled sizes are equal to % $L_{1,0} = 
$0.000$, $0.512$, $0.826$, and $1.000$ and 
the increments are equal to $\Delta L_{1,0} = 0.000$, $\Delta L_{1,1} = 0.512$, $\Delta L_{1,2} = 0.314$, and $\Delta L_{1,3} = 
0.174$. 
The information increments should sum to $1$.
This step process 
manifests as 
a gradual jump in the observed price as we saw in the methodology section. 

\clearpage

\section{The vanilla rearrangement algorithm}
\label{ssecRA}

Suppose that we want to rearrange the jump-event matrix of size $h \times q$. 
The rearrangement algorithm of \citet{puccetti2012computation} and \citet{embrechts2013model} loops over each column of a matrix to order it oppositely to the sum of the other columns.  
If the matrix has a fixed target in the last column \citep[see \textit{e.g.},][]{bernard2018rearrangement,bernard2017value}, the algorithm is as follows: 
\begin{enumerate}
	
	\item Randomly shuffle the elements (in each of the first $q - 1$ columns) to obtain the starting matrix of the algorithm. 
	The random shuffle 
	flattens (\textit{i.e.}, reduce the variability of) the row-sums.
	\begin{align*}
		{J}_n 
		= 
		\left[
		\begin{matrix*}[c]
			\phantom{.} \\
			\phantom{.} \\
			\phantom{.} \\
			\phantom{.} \\
			\phantom{.} \\
		\end{matrix*}
		\right.
		\underbrace{
			\begin{matrix*}[r]
				0.000 								& 0.000 						& 0.000  \\
				0.000 								& 0.000 						& 0.000  \\
				0.000 								& 0.000 						& 0.000  \\
				\underline{0.210} 		& 0.000 						& \underline{0.400}  \\
				0.000 								& \underline{0.217} & 0.000  \\
			\end{matrix*}
		}_{\substack{\text{Weighted} \\ \text{discontinuous} \\ \text{stock returns}}}
		% white space. the minuses are too close
		\begin{matrix*}[c]
			\phantom{.} \\
			\phantom{.} \\
			\phantom{.} \\
			\phantom{.} \\
			\phantom{.} \\
		\end{matrix*}
		\underbrace{
			\begin{matrix*}[c]
				-0.002 \\
				\phantom{-}0.003 \\
				-0.807 \\
				\phantom{-}0.004 \\
				-0.028 \\
			\end{matrix*}
		}_{\text{Target}}
		\left.
		\begin{matrix*}[c]
			\phantom{.} \\
			\phantom{.} \\
			\phantom{.} \\
			\phantom{.} \\
			\phantom{.} \\
		\end{matrix*}
		\right]
		%%%%%%
		&\xrightarrow[\text{Shuffle}]{\text{Random}}
		%%%%%%
		\left[
		\begin{matrix*}[c]
			\phantom{.} \\
			\phantom{.} \\
			\phantom{.} \\
			\phantom{.} \\
			\phantom{.} \\
		\end{matrix*}
		\right.
		\underbrace{
			\begin{matrix*}[r]
				0.000 								& 0.000 						& 0.000  \\
				\underline{0.210} 		& 0.000 						& 0.000  \\
				0.000 								& 0.000							& \underline{0.400}  \\
				0.000									& \underline{0.217} & 0.000 						  \\
				0.000 								& 0.000						  & 0.000  \\
			\end{matrix*}
		}_{\substack{\text{Rearranged} \\ \text{weighted} \\ \text{discontinuous} \\ \text{stock returns}}}
		% white space. the minuses are too close
		\begin{matrix*}[c]
			\phantom{.} \\
			\phantom{.} \\
			\phantom{.} \\
			\phantom{.} \\
			\phantom{.} \\
		\end{matrix*}
		\underbrace{
			\begin{matrix*}[c]
				-0.002 \\
				\phantom{-}0.003 \\
				-0.807 \\
				\phantom{-}0.004 \\
				-0.028 \\
			\end{matrix*}
		}_{\text{Target}}
		\left.
		\begin{matrix*}[c]
			\phantom{.} \\
			\phantom{.} \\
			\phantom{.} \\
			\phantom{.} \\
			\phantom{.} \\
		\end{matrix*}
		\right]
		\phantom{\rightarrow \cdots}
	\end{align*}
	%	\begin{align*}
	%		J_n^+ := \sum_{m = 1}^{q+1} (\gamma_{im})
	%		= 
	%		\begin{bmatrix*}[c]
	%			-0.002 \\ 
	%			\phantom{-}0.003  \\ 
	%			-0.807 \\ 
	%			\phantom{-}0.614  \\
	%			\phantom{-}0.189 
	%		\end{bmatrix*}
	%		&\xrightarrow[\text{Shuffle}]{\text{Random}}
	%		\begin{bmatrix*}[c]
	%			-0.002 \\ 
	%			\phantom{-}0.213  \\ 
	%			-0.406 \\ 
	%			\phantom{-}0.221  \\
	%			-0.028
	%		\end{bmatrix*}
	%		= \sum_{m = 1}^{q+1} (\gamma^\pi_{im}) := J_n^{\pi,+}
	%	\end{align*}
	
	\item 
	Iteratively rearrange the $l$-th column of the rearranged matrix $J^\pi_n$ so that it becomes oppositely ordered to the sum of the other columns, for $l = 1, ..., q - 1$. We never rearrange the target column, $l = q$.
	For $l = 1$, the algorithm immediately matches the stock jump with the ETF jump: 
	\begin{align*}
		% \cdots \rightarrow
		\phantom{{J}_n 
			= }
		\left[
		\begin{matrix*}[c]
			\phantom{.} \\
			\phantom{.} \\
			\phantom{.} \\
			\phantom{.} \\
			\phantom{.} \\
		\end{matrix*}
		\right.
		\underbrace{
			\begin{matrix*}[r]
				0.000 								& 0.000 						& 0.000  \\
				\underline{0.210} 		& 0.000 						& 0.000  \\
				0.000 								& 0.000							& \underline{0.400}  \\
				0.000									& \underline{0.217} & 0.000 						  \\
				0.000 								& 0.000						  & 0.000  \\
			\end{matrix*}
		}_{\substack{\text{Rearranged} \\ \text{weighted} \\ \text{discontinuous} \\ \text{stock returns}}}
		% white space. the minuses are too close
		\begin{matrix*}[c]
			\phantom{.} \\
			\phantom{.} \\
			\phantom{.} \\
			\phantom{.} \\
			\phantom{.} \\
		\end{matrix*}
		\underbrace{
			\begin{matrix*}[c]
				-0.002 \\
				\phantom{-}0.003 \\
				-0.807 \\
				\phantom{-}0.004 \\
				-0.028 \\
			\end{matrix*}
		}_{\text{Target}}
		\left.
		\begin{matrix*}[c]
			\phantom{.} \\
			\phantom{.} \\
			\phantom{.} \\
			\phantom{.} \\
			\phantom{.} \\
		\end{matrix*}
		\right]
		&\xrightarrow[\text{order}]{\text{Oppositely}}
		\left[
		\begin{matrix*}[c]
			\phantom{.} \\
			\phantom{.} \\
			\phantom{.} \\
			\phantom{.} \\
			\phantom{.} \\
		\end{matrix*}
		\right.
		\underbrace{
			\begin{matrix*}[r]
				0.000 								& 0.000 						& 0.000  \\
				0.000 								& 0.000 						& 0.000  \\
				\underline{0.210} 		& 0.000							& \underline{0.400}  \\
				0.000									& \underline{0.217} & 0.000 						  \\
				0.000 								& 0.000						  & 0.000  \\
			\end{matrix*}
		}_{\substack{\text{Rearranged} \\ \text{weighted} \\ \text{discontinuous} \\ \text{stock returns}}}
		% white space. the minuses are too close
		\begin{matrix*}[c]
			\phantom{.} \\
			\phantom{.} \\
			\phantom{.} \\
			\phantom{.} \\
			\phantom{.} \\
		\end{matrix*}
		\underbrace{
			\begin{matrix*}[c]
				-0.002 \\
				\phantom{-}0.003 \\
				-0.807 \\
				\phantom{-}0.004 \\
				-0.028 \\
			\end{matrix*}
		}_{\text{Target}}
		\left.
		\begin{matrix*}[c]
			\phantom{.} \\
			\phantom{.} \\
			\phantom{.} \\
			\phantom{.} \\
			\phantom{.} \\
		\end{matrix*}
		\right]
		% \rightarrow \cdots
		\phantom{ = J_n^\pi }
	\end{align*}
	
	\item 
	Repeat Step 2 until no further changes occur. That is, until a matrix $J_n^\pi$ is found with each column oppositely ordered to the sum of the other columns. The matrix will have row-wise sums with minimal variance. 
	
\end{enumerate}

\clearpage

%%%%%%%%%%%%%%%%%%%%%%%%%%%%%%%%%%%%%%%%%%%%%%%%%%%%%%%%%%%%%%%%%%%%%%%%%%%%%%%%%%%%%%%%%%%%%%%%%%%%%%%%%%%%%%%%%%%%%%%%%

%%%%%%%%%%%%%%%%%%%%%%%%%%%%%%%%%%%%%%%%%%%%%%%%%%%%%%%%%%%%%%%%%%%%%%%%%%%%%%%%%%%%%%%%%%%%%%%%%%%%%%%%%%%%%%%%%%%%%%%%%

\end{document}